\documentclass[12pt,a4paper]{article}            
 \usepackage[skins,theorems]{tcolorbox}
\tcbset{highlight math style={enhanced,
  colframe=red,colback=white,arc=0pt,boxrule=1pt}}
  \usepackage[bookmarksopen, bookmarksnumbered, bookmarksopenlevel=2]{hyperref}
  \usepackage{tikz}
  \usepackage{tikz-3dplot}
  \usepackage{multirow}
 \usetikzlibrary{calc}
 \usetikzlibrary{decorations} %
 \usepackage[UKenglish]{babel}
 \usepackage[toc,page]{appendix}
 \usepackage{amsmath}
 \usepackage{amssymb}
 \usepackage{graphicx}
 \usepackage{mathrsfs}
 \usepackage{hhline}
 \usepackage[bf]{caption}
\usepackage{cite}
\usepackage[vcentermath]{youngtab}
\usepackage{geometry}
\usepackage{slashed}
\usepackage{color}
\usepackage{stackrel}
\usepackage{tikz-cd} 
\usepackage{cancel} 
\usepackage[normalem]{ulem}
\usepackage{empheq}
\usepackage{arydshln}

\newenvironment{eqn*}{\begin{equation*}\begin{aligned}}{\end{aligned}\end{equation*}\noindent}

\newcommand{\bqa}{\begin{eqnarray}}
\newcommand{\eqa}{\end{eqnarray}}

\hypersetup{
    pdftitle={},
    pdfauthor={},
    pdfsubject={}
}
\numberwithin{equation}{section}
\numberwithin{table}{section}

\makeatletter

\newcommand{\be}{\begin{equation}}
\newcommand{\ee}{\end{equation}}
\newcommand{\beq}{\begin{equation}}
\newcommand{\eeq}{\end{equation}}
\newcommand{\ba}{\begin{aligned}}
\newcommand{\ea}{\end{aligned}}

\newcommand{\bea}{\begin{eqnarray}}
\newcommand{\eea}{\end{eqnarray}}

\newcommand{\cT}{\mathcal{T}}

\newcommand{\cL}{\mathcal{L}}

\newcommand{\cN}{\mathcal{N}}

\newcommand{\cA}{\mathcal{A}}

\newcommand{\cF}{\mathcal{F}}
\newcommand{\cI}{\mathcal{I}}
\newcommand{\cJ}{\mathcal{J}}

\newcommand{\cS}{\mathcal{S}}

\newcommand{\cV}{\mathcal{V}}

\newcommand{\cM}{\mathcal M}

\newcommand\bi{\begin{itemize}}
\newcommand\ei{\end{itemize}}

\def\Tr{\mathop{\mathrm{Tr}}\nolimits}
\def\tr{\mathop{\mathrm{tr}}\nolimits}

\def\unit{{1\kern-.65ex {\rm l}}}
\def\1{{1\kern-.65ex {\rm l}}}

\def\CU{{\cal U}}

\def\CX{{\cal X}}

\def\CZ{{\cal Z}}

\def\ii{{\rm i}}

\newcount\hour \newcount\minute
\hour=\time \divide \hour by 60
\minute=\time
\def\now{%
\ifnum \hour<13
  \ifnum \hour=0 \advance \hour by 12 \number\hour:\else \number\hour:\fi%
     \ifnum \minute<10 0\fi%
     \number\minute%
\ A.M.%
\else \advance \hour by -12 \number\hour:%
  \ifnum \minute<10 0\fi%
  \number\minute%
  \ P.M.%
\fi%
}

\def\fnote#1#2{\begingroup\def\thefootnote{#1}\footnote{#2}
     \addtocounter{footnote}{-1}\endgroup}

\makeatother

\begin{document}

\begin{flushright}
{\tt\normalsize UPR-1330-T}\\
\end{flushright}

\vskip 40 pt
\begin{center}
{\large \bf
  Non-perturbative Resolution of Strong Coupling\vspace{5pt}\\ Singularities in 4d N=1 Heterotic/M-theory
} 

\vskip 11 mm

Mirjam Cveti\v{c}${}^{1,2,3}$ and Max Wiesner${}^{4}$

\vskip 11 mm
${}^1$ {\it Department of Physics and Astronomy, University of Pennsylvania, Philadelphia, PA 19104, USA}  \\[3 mm]
${}^2$ {\it Department of Mathematics, University of Pennsylvania, Philadelphia, PA 19104, USA}  \\[3 mm]
${}^3$ {\it Center for Applied Mathematics and Theoretical Physics, University of Maribor, Maribor, Slovenia} \\[3 mm]
${}^4${\it Jefferson Physical Laboratory, Harvard University, Cambridge, MA 02138, USA}  \\[3 mm]


\fnote{}{cvetic at physics.upenn.edu, mwiesner at fas.harvard.edu}

\end{center}


\begin{abstract}
We investigate the interior of the moduli space of four-dimensional $\cN=1$ theories of gravity arising from compactifications of the $E_8\times E_8$ heterotic string on Calabi--Yau threefolds. By studying the threshold corrections to the coupling of the heterotic gauge groups, we infer the existence of a strong coupling singularity for one of the perturbative heterotic gauge groups, which effectively yields an additional finite distance boundary of the classical scalar field space. In heterotic M-theory, this boundary maps to a domain wall solution for which the gauge coupling and the warp factor on one of the Horava--Witten 9-branes diverge, thus highlighting the gravitational origin of the classical strong coupling singularity. The divergence of the warp factor is, however, regulated once non-perturbative effects are taken into account, as we demonstrate by studying the instanton corrections to the 5d BPS domain wall equations. This regularization implies that the classical strong coupling boundary of the scalar field 4d $\cN=1$ heterotic/M-theory is resolved, indicating that, at the quantum level, the field space can be extended beyond this classical boundary.
 \end{abstract}

\vfill

\thispagestyle{empty}
\setcounter{page}{0}
 \newpage

\tableofcontents
\setcounter{page}{1}
\section{Introduction}
Scalar field spaces play a crucial role in theories of gravity since all parameters of such theories are expected to be determined by the vacuum expectation values of scalar fields. The properties and structure of scalar field spaces in gravity hence reveal important information about the features of the underlying gravitational theory, such as the possible values of couplings or masses of states. For this reason, scalar field spaces are a central object of study for the Swampland program~\cite{Vafa:2005ui} which prominently features them in many of its core statements~\cite{Ooguri:2006in}, see also \cite{Palti:2019pca,vanBeest:2021lhn, Agmon:2022thq} for reviews. To make concrete statements, the focus often lies on scalar field spaces of highly supersymmetric theories with eight or more supercharges. The reason is that supersymmetry, on the one hand, ensures that the scalar fields are actual moduli of the theory, i.e., no potential is generated for them. On the other hand, at least for certain sub-sectors of the theory, enhanced supersymmetry protects the moduli space from quantum corrections such that the classical moduli space metric obtained from supergravity is exact. 

The situation is considerably different in four-dimensional theories with minimal supersymmetry. In such 4d $\cN=1$ theories, all scalar fields reside in chiral multiplets for which a non-trivial potential can and, for genuine $\cN=1$ theories, is expected to be generically generated~\cite{Palti:2020qlc}. The scalar field spaces of these theories are thus not actual moduli spaces of the theory. Still, if the potential is sufficiently small, it can effectively be treated as a quasi-moduli space. In this work, we consider explicit 4d $\cN=1$ effective theories of gravity obtained from compactifications of heterotic string theory on Calabi--Yau threefolds. The scalar field space of these theories is spanned by the four-dimensional dilaton, the geometric moduli of the Calabi--Yau threefold, and bundle moduli. For most of this work, we exclusively focus on the dilaton and the geometric moduli by considering the special case of heterotic compactifications with so-called standard embedding. The geometry of the corresponding quasi-moduli space can be described in terms of a K\"ahler potential, which, in the weak coupling limit for the heterotic string, is determined purely by the geometry of the compact geometry.
On the other hand, away from the strict weak-coupling limit, the K\"ahler potential receives corrections both at the perturbative and non-perturbative level. Since the K\"ahler potential of 4d $\cN=1$ theories does not enjoy any protection these corrections are notoriously difficult to compute directly. As a consequence, unlike for 4d theories with enhanced supersymmetry, the interior of the quasi-moduli space of heterotic Calabi--Yau threefolds remains largely unexplored. This paper aims to improve this situation and obtain insights into the non-perturbative structure of the quasi-moduli space of such 4d $\cN=1$ theories. 

To achieve this goal, we do not attempt to directly compute the corrections to the K\"ahler potential of the full $\cN=1$ theory. Instead, our strategy is to infer the structure of the field space from studying the physical parameters of the theory. It is well-known from theories with extended supersymmetry that the non-perturbative structure of the scalar field space is reflected in the parameters of the theory, for example, as strong coupling regimes for gauge theories or additional light states in the theory. The classic example for this is the physical explanation of the conifold singularity of Calabi--Yau threefolds in terms of light hypermultiplets in Type II string theory~\cite{Strominger:1995cz}. In this work, we focus on the heterotic $E_8\times E_8$ string theory with standard embedding and use the gauge coupling of the unbroken $E_8$ factor as the parameter that is sensitive to the field space structure of the underlying $\cN=1$ theory. At tree level, the heterotic gauge coupling is simply given by the heterotic 4d dilaton. Still, at one-loop, the gauge couplings receive threshold corrections, which have been computed in~\cite{Kaplunovsky:1987rp,Dixon:1990pc}. As a result of these threshold corrections, the gauge coupling of the hidden $E_8$ diverges along a real co-dimension one locus in the field space, indicating the existence of a strong coupling singularity of the heterotic string compactified on Calabi--Yau threefolds. Strong coupling singularities in 4d $\cN=1$ gauge theories obtained from string theory have previously been discussed in~\cite{Atiyah:2000zz,Acharya:2000gb}. In these works, the strong coupling phase of gauge theories on space-time filling D6-branes of Type IIA string theory wrapping three-cycles in non-compact Calabi--Yau threefolds have been analyzed and interpreted as geometric transitions in the uplift to M-theory compactified on non-compact $G_2$-manifolds. In contrast to these strong coupling effects for \emph{open string} gauge theories, here, we are dealing with strong coupling singularities in the \emph{closed string} sector which cannot be decoupled from gravity. 

The perturbative heterotic analysis based on the threshold corrections suggests that the scalar field space ends at the strong coupling singularity. However, this conclusion cannot hold at the non-perturbative level since the scalar field spaces of 4d $\cN=1$ theories are complex, and complex spaces only have singularities of at least complex co-dimension one. Therefore, at the quantum level, it has to be possible to circumvent the strong coupling singularity by suitably tuning axionic fields such that the scalar field space extends beyond the strong coupling singularity. Indeed, in non-gravitational 4d $\cN=1$ theories, it is known that the moduli space extends beyond the strong coupling singularity as was demonstrated in~\cite{Atiyah:2000zz,Acharya:2000gb} by explicitly constructing the confined phase of the D6-brane gauge theory. However, a quantitative derivation of the resolution of the classical singularity through M2-/D2-brane instantons has not been found yet. 

In this paper, we demonstrate how instanton effects resolve the strong coupling singularity of the perturbative heterotic string. Since the singularity arises in the closed string sector, its resolution is similarly gravitational in nature. The non-perturbative corrections responsible for this resolution can be computed explicitly via duality to heterotic M-theory, where they arise as corrections to the equations of motion for domain walls that saturate the Bogomol'nyi–Prasad–Sommerfield (BPS) bound. As is well-known, the heterotic string on a Calabi--Yau threefold $X$ is dual to M-theory on $X\times S^1/\mathbb{Z}_2$ with two 9-branes located at the orbifold fixed points~\cite{Horava:1995qa}. For standard embedding, both 9-branes have induced 5-brane charge proportional to the integrated second Chern class of $X_3$ \cite{Lukas:1998yy}. As a consequence, the volume of $X$ gets a non-trivial profile along the interval such that, effectively, we obtain a five-dimensional version of a dilatonic domain wall~\cite{Cvetic:1992bf,Cvetic:1994ya,Cvetic:1996vr,Skenderis:1999mm}. In the five-dimensional parent theory, the volume of $X$ is a scalar in the universal hypermultiplet. Using the classical metric for the universal hypermultiplet in the BPS equations of motion~\cite{Cvetic:1992bf,Cvetic:1994ya,Cvetic:1996vr,Skenderis:1999mm} one realizes that the gauge coupling on the hidden domain wall diverges for some finite length of the interval reproducing the perturbative strong coupling singularity of the heterotic string at a finite distance in the field space. For the classical moduli space metric, the BPS equations further imply that the warp factor of the spacetime metric diverges as well for this size of the interval. The divergence of the warps factor indicates that spacetime ends at the location of the hidden 9-brane. This, in turn, implies that also the moduli space of the 4d $\cN=1$ theory obtained by reducing the domain wall solution on the interval ends at the strong coupling singularity since the length of the interval cannot be made arbitrarily large.  

An important benefit of this dual M-theory description is that certain non-perturbative corrections can be taken into account explicitly. The relevant corrections arise as instanton corrections to the universal hypermultiplet of the 5d parent theory and enter the domain solution through the dependence of the BPS equations on the 5d moduli space metric. The fact that the non-perturbative effects manifest as corrections to the BPS domain wall equations illustrates that the resolution of the strong coupling singularity in the closed string sector is indeed of gravitational nature. The hypermultiplet moduli space of the five-dimensional parent theory receives corrections both at the perturbative and non-perturbative level. Of particular interest are the instanton corrections to the hypermultiplet moduli space, which arise from Euclidean M2- and M5-branes wrapping cycles in the Calabi--Yau threefold. Via duality, these instanton effects can be identified with D2-brane and NS5-brane instanton effects on the hypermultiplet moduli space of Calabi--Yau threefold compactifications of Type IIA string theory. In the case of D2-brane instanton, these effects have been computed via duality in~\cite{RoblesLlana:2006ez,RoblesLlana:2006is,RoblesLlana:2007ae, Alexandrov:2008gh,Alexandrov:2009zh,Alexandrov:2012au,Alexandrov:2017qhn} and are by now well-understood.\footnote{In the context of the distance conjecture~\cite{Ooguri:2006in}, these kind of instantons have been analyzed in the mirror dual Type IIB setup in \cite{Marchesano:2019ifh,Baume:2019sry}, where it was shown that they obstruct certain classical infinite distance limits which otherwise violate the emergent string conjecture~\cite{Lee:2019oct}; see also \cite{Alvarez-Garcia:2021pxo} for a similar study in M-theory compactifications on Calabi--Yau threefolds.}  For our purpose, the relevant corrections originate from the 5-brane instantons, whose computation is more involved. Still, progress on the computation of 5-brane instanton corrections has been achieved in~\cite{Alexandrov:2006hx,Alexandrov:2009vj,Alexandrov:2010ca,Alexandrov:2014mfa,Alexandrov:2014rca,Alexandrov:2023hiv} which we apply to our setup in this work.

Since the hypermultiplet moduli space metric enters the 5d domain wall equations, the 5-brane instanton corrections to the hypermultiplet metric indirectly also correct the domain wall solution of heterotic M-theory.  Using the results of \cite{Alexandrov:2006hx,Alexandrov:2009vj,Alexandrov:2010ca,Alexandrov:2014mfa,Alexandrov:2014rca,Alexandrov:2023hiv} allows us to capture the instanton corrections on these domain walls. The central result is that, due to the quantum corrections to the universal hypermultiplet metric, the warp factor of the domain solution remains finite even if the coupling of the gauge theory on the hidden domain wall diverges. Thus, at the quantum level, the strong coupling behavior does not imply that spacetime ends at the location of the strongly coupled 9-brane such that also the scalar field space of the effective 4d $\cN=1$ theory extends beyond this regime. Our results hence imply that non-perturbative corrections enable us to move beyond the classical real co-dimension one boundary, indicating that, indeed, the heterotic strong coupling singularity is resolved into a singularity of at most complex co-dimension one at the quantum level.  This is the main result of this paper and provides an explicit confirmation that non-perturbative corrections in 4d $\cN=1$ theories resolve the strong coupling singularities in the closed string sector. \\

The rest of this paper is structured as follows: In Section~\ref{sec:perthet} we review the threshold corrections to the gauge coupling of the heterotic string compactified on Calabi--Yau threefolds and show that, in the simplest cases, they lead to strongly coupled gauge theories at finite distance in the moduli space. We then review the dual M-theory setup of~\cite{Lukas:1998yy} in Section~\ref{sec:classDW} and use the heterotic/M-theory dictionary to show that the classical domain wall solution reproduces the strong coupling singularity of the perturbative heterotic string. In Section~\ref{sec:nonpert}, we then discuss the effect of non-perturbative corrections on the domain wall solution, which is mainly based on the recent results of \cite{Alexandrov:2023hiv}. After reviewing their main results, we apply these to the domain solution to show that the strong coupling singularity is resolved at the quantum level. We present our conclusions and possible future directions in section~\ref{sec:concl}. 

\section{Threshold corrections in Heterotic String Theory}\label{sec:perthet}
Consider a simple compactification of the heterotic $E_8\times E_8$ string on a Calabi--Yau threefold $X_3$. Classically, this theory has a moduli space spanned by the dilaton of the four-dimensional effective action, the complex structure moduli of $X_3$, and the complexified K\"ahler moduli of $X_3$ 
\begin{equation}\label{def:moduli}
 T_i = \frac{1}{l_s^2} \int_{C^i} (\omega +iB_2) \,,
\end{equation} 
where $\omega$ is K\"ahler form on $X_3$, $\{C^i\}$, $i=1,\dots, h^{1,1}(X_3)$ a basis of generators of the effective cone of curves, $l_s$ the heterotic string length and $B_2$ the heterotic $B$-field. In addition, there are moduli associated with the choice of gauge bundle in the two $E_8$ gauge factors. Let us denote by $V_a$, $a=1,2$, the gauge bundle in the $a$-th $E_8$ factor. In this work, we mostly focus on the case of standard embedding for which
\begin{equation}\label{2,2locus}
 V_1 = T X_3 \,,\qquad V_2 = \mathcal{O}_{X_3}\,,
\end{equation}
where $TX_3$ denotes the tangent bundle of $X_3$. This choice of gauge bundle satisfies the Bianchi identity 
\begin{equation}\label{het:Bianchi}
 dH_3 = \frac{\alpha'}{2} \left(\tr F^{(1)} \wedge F^{(1)}  + \tr F^{(2)} \wedge F^{(2)}  -{\rm tr} R^2  \right)\,,
\end{equation}
without the need to introduce space-time filling NS5-branes. In the above expression $F^{(a)}$ is the gauge field strength associated to the bundle $V_a$ and $H_3$ is the field strength of the heterotic B-field. The choice of bundles in \eqref{2,2locus} breaks the first $E_8$ factor to $E_6$ while keeping the second $E_8$ factor intact. In principle, we can consider deformations of the first bundle while keeping $c_2(V_1) =c_2(X_3)$. From the perspective of the heterotic string worldsheet, these deformations break the $(2,2)$ worldsheet supersymmetry to $(0,2)$. Still, for most parts of this work, we do not consider such deformations but stick to the choice \eqref{2,2locus}, which we refer to as '(2,2)-locus' in the bundle moduli space.

In addition to the geometric moduli, there is the heterotic dilaton $S_{\rm het}$ that determines the effective string coupling $g_4$ of the four-dimensional theory via
\begin{equation}
\frac{1}{g_4^2} = \langle S_{\rm het}+\bar{S}_{\rm het}\rangle\,. 
\end{equation}
The subset of the full moduli space spanned by $S_{\rm het}$ and $T_i$ is described by a K\"ahler potential which, at string tree-level, reads
\begin{equation}\label{Kpothet}
 K = -\log(S_{\rm het}+\bar{S}_{\rm het}) -\log \left[\frac16 \int_{X_3} J^3\right]\,, 
\end{equation} 
where the second term should be viewed as a function of the real parts of the $T_i$ in \eqref{def:moduli}. The K\"ahler potential determines the kinetic terms for these moduli. On the other hand, the kinetic terms of the gauge fields associated with the $a$-th gauge group are determined by the gauge kinetic function $f_a$, which, at string tree level, is given by
\begin{equation}\label{treelevelgaugekinetic}
 f_a = k_a S \,,
\end{equation}
where $k_a$ is the level of the Kac-Moody algebra on the worldsheet. Accordingly, the gauge coupling of the heterotic gauge group is, at tree level, given by $\text{Re}\, S_{\rm het}$. 

However, one-loop threshold effects correct the gauge coupling of the $a$-th gauge factor as 
\begin{equation}
\frac{16\pi^2}{g_a^2} = k_a \text{Re}\,S_{\rm het} + b_a \log \frac{M_s}{\mu^2} +\Delta_a\,,
\end{equation}
where $\mu$ is some renormalization scale, $b_a$ the one-loop $\beta$-function coefficient of the $a$-th gauge factor $\beta_a = b_a g_a^3/(16\pi^2)$ and $\Delta_a$ encodes the string-theoretic one-loop threshold corrections. These have been computed in \cite{Kaplunovsky:1987rp,Dixon:1990pc} and take the general form 
\begin{equation}\label{def:Deltaa}
\Delta_a = \frac{1}{16\pi^2}\int_\Gamma \frac{{\rm d}^2\tau }{\tau_2}\left( \mathcal{B}_a(\tau, \bar \tau)-b_a\right)\,. 
\end{equation}
Here $\Gamma$ is the fundamental domain of the worldsheet $T^2$ with complex structure $\tau$ and $\mathcal{B}_a$ can be expressed as 
\begin{equation}
\mathcal{B}_a(\tau,\bar{\tau}) =|\eta(\tau)|^{-4}\cdot \sum_{\text{even} \,\mathbf{s}}(-1)^{s_1+s_2}\frac{\mathrm{d}Z_{\Psi}(s,\bar\tau)}{(2\pi i) {\rm d}\bar\tau}\cdot \Tr_{s_1}\left(Q^2(-1)^{s_2F}q^H\bar{q}^{\bar H}\right)\,.
\end{equation}
In this expression, the sum runs over $s_i=0 (1)$, labeling Neveu-Schwarz (Ramond) boundary conditions on the $(2,2)$ worldsheet, and the trace is over the internal Calabi--Yau conformal field theory (CFT). Furthermore, $Z_\Psi$ is the partition function for worldsheet fermions associated with the four extended directions. Of particular interest to us is the trace over the internal part of the CFT since this encodes the dependence of the threshold corrections on the moduli of the internal CFT. 

In simple toroidal setups, these thresholds can be explicitly computed. We therefore consider now the special case $X_3=T^6/G$ for some of the orbifold group $G$. Moreover we focus only on the $\cN=2$ subsector of the K\"ahler moduli $T_i$, $i=1,2,3$, corresponding to the (complexified) volumes of $T^6=(T^2)^3$. For this setup, the threshold corrections have been computed explicitly in \cite{Dixon:1990pc}. In this case $\mathcal{B}_a$ simplifies to 
\begin{equation}\label{Baintegral}
\mathcal{B}_a = \hat{Z}_{\rm torus}(\tau, \bar{\tau}) \mathcal{C}_a(\tau) \,,
\end{equation} 
with the torus-partition function
\begin{equation}
\hat{Z}_{\rm torus}(\tau, \bar{\tau})\equiv \sum_{(p_L,p_R)\in \Gamma_{2,2}}q^{p_L^2/2}\bar{q}^{p_R^2/2}\,.
\end{equation} 
It can be shown that the function $\mathcal{C}_a$ is constant such that $\Delta_a$ for toroidal orbifolds satisfies
\begin{equation}
 \Delta_a = b_a \int\frac{\mathrm{d}^2\tau}{\tau_2}\left(\hat{Z}_{\rm torus}(\tau,\bar{\tau}) -1 \right)\,.
\end{equation}
The remaining integral can be explicitly evaluated to give
\begin{equation}
\Delta_a =- b_a \prod_{i=1}^3 \log[\text{Im}\,T_i|\eta(iT_i)|^4] +{\rm const}\,.
\end{equation}
with $\eta$ the Dedekind function defined as 
\begin{equation}\label{def:etafunc}
\eta(\tau) = e^{\frac{\pi i \tau}{12}}\prod_{n=1}^\infty \left(1-e^{2\pi i n \tau} \right)\,. 
\end{equation} 
With this preparation, we can now discuss the effect of threshold corrections to the gauge couplings.

\subsection{Strong Coupling for the Heterotic String}
First, let us simplify our setup further by setting all three moduli equal $T_1=T_2=T_3\equiv T$ and focusing on the overall modulus $T$.  In this case the correction to the coupling of the unbroken $E_8$ factor is given by 
\begin{equation}
 \frac{1}{g_{E_8}^2} =  k_2 \,\text{Re}\,S_{\rm het} - \frac{3 b_{E_8}}{16\pi^3} \log [\text{Im}\,T |\eta(iT)|^4]\,,
\end{equation} 
where for the unbroken $E_8$ the $\beta$-function coefficient is negative (its exact value depends on the chosen orbifold group $G$). We can now use \eqref{def:etafunc} to expand the above expression for large $\text{Re}\,T$ as 
\begin{equation}\label{gE8}
  \frac{1}{g_{E_8}^2} =  k_2\, \text{Re}\,S_{\rm het} - \frac{ |b_{E_8}|}{16\pi^3} \text{Re}\,T + \mathcal{O}(\log\text{Im}\,T)\,, 
\end{equation} 
where we used $b_{E_8}<0$. From the above expression, it is apparent that there exists a value $0<\text{Re}\,T_{\rm max}<\infty$ at which the unbroken $E_8$ gauge group becomes infinitely strongly coupled. More precisely, this strong coupling behavior occurs for
\begin{equation}
 \text{Re}\,T_{\rm max} \simeq \frac{16 \pi^2 k_2}{|b_{E_8}|}\text{Re}\,S_{\rm het}\,,
\end{equation}
where we have dropped the subleading terms in \eqref{gE8}. Thus, the part of the moduli space spanned by $(S,T)$ in which we can describe the theory as a perturbative heterotic string is constrained to the regime $\text{Re}\,T<\text{Re}\,T_{\rm max}$ and the correct description of the theory beyond that regime is unclear.

Let us note that also at the level of the superpotential and scalar potential, it can be seen that the perturbative heterotic description breaks down around $\text{Re}\,T=\text{Re}\,T_{\rm max}$. Given the negative $\beta$-function, we notice that there is a non-perturbative contribution to the superpotential due to gaugino condensation
\begin{equation}
 W = M_{\rm het}^3 \exp\left(\frac{f_a(S,T)}{2 b_a}\right)\,.
\end{equation} 
where $f_a(S,T)$ is the gauge kinetic function of the $E_8$-gauge theory. Since the gauge kinetic function is moduli-dependent, the superpotential in principle also depends on both moduli. For the case $X_3=T^6/G$ the threshold corrections to the holomorphic gauge kinetic functions can be written as
\begin{equation}
f_a = k_a S - \frac{b_a}{16\pi^3} \log \eta(iT)^{12} \,,
\end{equation} 
where $T$ is again the overall K\"ahler modulus of $T^6$. As a consequence, also the superpotential has a non-trivial $T$-dependence (see \cite{Cvetic:1991qm} for a detailed discussion on this)
\begin{equation}\label{Weta}
 W = M_{\rm het} ^3 \frac{e^{\frac{k_a S}{2 b_a}}}{\eta(iT)^6} \,. 
\end{equation}
In general the $SL(2,\mathbb{Z})$ duality group acting on $T$ as 
\begin{equation}
 T \rightarrow \frac{a T+ b}{cT+d} \,,\qquad \text{for}\qquad \begin{pmatrix} a & b\\c&d  \end{pmatrix} \in SL(2,\mathbb{Z})\,,
\end{equation} 
requires $W$ to be a modular form of weight-3 such that the combination 
\begin{equation}
 e^G \equiv e^{K/2} W \,,
\end{equation} 
is modular invariant. Since $\eta$ is a weight-$1/2$ modular form, the superpotential in \eqref{Weta} is indeed a weight-3 modular form. Still, modular invariance allows to multiply \eqref{Weta} by a modular invariant function $H(T)$ such that 
\begin{equation}\label{Wfinal}
 W = \frac{\Omega(S) H(iT)}{\eta^6(iT)}\,,
\end{equation}
where $\Omega(S)$ is in principle some general function of $S$ which in the above case is simply given by $\Omega(S) = e^{\frac{k_a S}{2b_a}}$. The general form of $H(T)$ was derived in \cite{Cvetic:1991qm} based on the assumption that the only poles of $W$ should be at $T\rightarrow \infty$. In this case $H(T)$ takes takes the form 
\begin{equation}
H(iT) = \left(\frac{G_4(iT)}{\eta^8(iT)}\right)^m \left(\frac{G_6(iT)}{\eta^{12}(iT)}\right)^n \mathcal{P}(j(iT))\,,
\end{equation} 
where $G_{4}$ ($G_6$) is the holomorphic weight-4 (-6) Eisenstein series and $\mathcal{P}$ is a polynomial in the $j$-invariant. The special values of the $j$-invariant are given by 
\begin{equation}
j(\ii \infty) = \infty\,, \qquad j(\ii ) =1 \,,\qquad j(e^{2\pi i/3}) = 0\,. 
\end{equation} 
Let us note that we cannot replace $\mathcal{P}(j(iT))$ by a rational function of $j$ that vanishes for $T\rightarrow \infty$ without introducing poles at finite values of $T$. As a consequence, for $\text{Re}\,T\rightarrow \infty$ the superpotential diverges and in particular we expect 
\begin{equation}
 |W| \sim \mathcal{O}(M_{\rm het}^3) \,,\quad \text{for}\quad \text{Re}\,T\simeq \text{Re}\,T_{\rm max}\,. 
\end{equation}
Thus, the strong coupling singularity for the unbroken $E_8$ group is accompanied by a non-perturbative superpotential of the order of the quantum gravity cut-off, which for $\text{Re}\,S_{\rm het}\gg 1$ is given by $M_{\rm het}$. Similarly, in 4d $\cN=1$ theories, the scalar potential is given in terms of the superpotential as
\begin{equation}
 V = e^K \left(g^{i \bar{j}} D_i W \bar{D}_{\bar j} \bar{W} - 3|W|^2 \right)\,,
\end{equation}
in Planck units. Here, $D_i = \partial_i + \partial_i K$ is the K\"ahler covariant derivative. At $\text{Re}\,T_{\rm max}$ the scalar potential is of order of the quantum gravity cutoff signaling that the underlying EFT, which in our case is the supergravity action derived from the perturbative heterotic string, becomes invalid \cite{Hebecker:2018vxz,Scalisi:2018eaz,vandeHeisteeg:2023uxj}. 

Let us stress that the analysis so far is just based on the leading perturbative heterotic threshold corrections and we need to supplement the above considerations by a non-perturbative analysis. Before we turn to this in the next section, let us notice that one can observe the strong coupling effect at finite $\text{Re}\,T$ also in cases other than $X_3=T^6/G$. We, therefore, notice that on the (2,2) locus in the bundle moduli space, the computation of the integral in \eqref{Baintegral} is, in fact, the same as the computation of the topological genus-one free energy, $F_1$, for Type II strings as performed \cite{Bershadsky:1993ta,Bershadsky:1993cx}.\footnote{In Type II compactifications, the genus-one free energy $F_1$ computes a higher-derivative correction to the effective action. As such, $F_1$ can be used to capture the moduli dependence of the quantum gravity cutoff (or species scale) in the vector multiplet moduli space of the effective $\cN=2$ four-dimensional theory as recently discussed in~\cite{vandeHeisteeg:2022btw,Cribiori:2022nke}. } Importantly, the computation of $F_1$ can also be carried out for general Calabi--Yau $X_3$. We are again interested in the limit in which the overall K\"ahler modulus of $X_3$, which we denote by $T$, becomes large. In this limit, the integral over the genus-one partition function is given by  \cite{Bershadsky:1993ta,Bershadsky:1993cx}
\begin{equation}
F_1 \rightarrow \frac{1}{12} \int_{X_3} c_2 \wedge \omega + \dots \,. 
\end{equation}
In general, for standard embedding on the (2,2) locus, the coefficient of the leading term of the one-loop threshold corrections linear in the moduli is then given by 
\begin{equation} 
 \Delta_a \rightarrow b_{E_8} \left(\frac{\gamma}{(\text{vol}(X_3))^{1/3}} \int_{X_3} c_2 \wedge \omega  \right) \text{Re}\,T +\dots\,,
\end{equation} 
where we absorbed all proportionality factors in the constant $\gamma$. The dots indicate all terms subleading in $\text{Re}\,T$. For the coupling of the hidden $E_8$, this means 
\begin{equation}\label{leadingcorrections}
\frac{1}{g_{E_8}^2} = \text{Re}\,S_{\rm het} - \frac{|b_{E_8}|}{ 16\pi ^3}\left(\frac{\gamma}{(\text{vol}(X_3))^{1/3}} \int_{X_3} c_2 \wedge \omega  \right) \text{Re}\,T +\dots\,. 
\end{equation}
We, therefore, expect also for perturbative heterotic compactifications on general Calabi--Yau threefold $X_3$ with standard embedding \eqref{2,2locus} that there is a maximal value $\text{Re}\,T_{\rm max}\sim \text{Re}\,S_{\rm het}$ where the unbroken $E_8$ gauge group becomes strongly coupled and the effective theory derived from the perturbative heterotic string theory breaks down. 

The summary of the perturbative analysis of this section is that for fixed heterotic string coupling $\text{Re}\,S_{\rm het}$ we hit a strong coupling singularity at finite $\text{Re}\,T=\text{Re}\,T_{\rm max}$. Based on this analysis, one might conclude that the scalar field space of the effective $\cN=1$ theory ends at the wall corresponding to the strong-coupling singularity along which $g_{E_8}=\infty$. Since along this boundary both $\text{Re}\,T$ and $\text{Re}\,S$ are finite, this boundary of the field space would be at finite distance in the field space metric derived from the K\"ahler potential in \eqref{Kpothet}. However, such a real co-dimension one singularity should not exist in the complex chiral multiplet field space of the effective $\cN=1$ theory. Instead, all singularities are at most of complex co-dimension one. To achieve this, non-perturbative corrections have to be taken into account that resolve the $\text{codim}_{\mathbb{R}}=1$ singularity into a $\text{codim}_{\mathbb{C}}=1$ singularity, see \cite{Atiyah:2000zz} for a similar argument for singularities in gauge theories decoupled from gravity geometrically engineered in M-theory compactified on non-compact $G_2$ manifolds. In order to compute such quantum effects, we need to consider dual setups. Here, we first discuss the strong coupling singularity from a dual F-theory perspective before we discuss the dual M-theory perspective in the next section.

\subsection{Duality to F-theory}
As a first dual description, we consider F-theory compactified on elliptically fibered Calabi--Yau fourfolds. We therefore recall that the heterotic string on an elliptically fibered Calabi--Yau threefold $T^2 \hookrightarrow X_3 \stackrel{p}{\rightarrow} B_2$ is dual to F-theory on an elliptically fibered Calabi--Yau fourfold $T^2 \hookrightarrow Z_4 \stackrel{\pi}{\rightarrow} B_3$ for which the base, $B_3$, is itself rationally fibered $\mathbb{P}^1\hookrightarrow B_3\stackrel{\rho}{\rightarrow} B_2$ over the same base as $X_3$~\cite{Morrison:1996na,Morrison:1996pp}.  In the simplest case, $X_3$ is a smooth Weierstrass model over $B_2$ for which the K\"ahler moduli of $X_3$ can be split as $\{T_0, T_a\}$, $a=1,\dots, h^{1,1}(B_2)$, where $\text{Re}\,T_0$ measures the volume of the elliptic fiber and $\text{Re}\,T_a$ the volumes of curves in $B_2$ in heterotic string units. Via heterotic/F-theory duality, these latter moduli are identified with the volume of divisors in $B_3$ that are vertical w.r.t. the projection $\rho$. More precisely, in the dual F-theory frame, we can define the chiral fields~\cite{Grimm:2010ks} 
\begin{equation}
 \cT^a = \frac{1}{l_{\rm IIB}^4}\int_{\rho^*(C_a)} \left(\frac12 J^2 + i C_4\right)\,,
\end{equation}
where $C_4$ is the Type IIB RR four-form and $l_{\rm IIB}$ is the Type IIB string length. Moreover, $C_a$ are the K\"ahler cone generators of $B_2$. Upon F-theory/heterotic duality, we then identify 
\begin{equation}
\text{Re}\,\cT^a = \eta^{ab}\, \text{Re}\,T_b \,,
\end{equation} 
where $\eta^{ab}$ is the intersection form on $B_2$. On the other hand, the modulus $T_0$ of the heterotic theory maps to a complex structure modulus of the $X_4$ controlling the stable degeneration limit of the rationally fibered $B_3\to B_2$. Finally, in the adiabatic limit, in which the base $B_2$ is very large compared to the fibers of $\rho$ and $p$, the four-dimensional heterotic dilaton can be identified with the volume of the sections of $B_3$ on the F-theory side of the duality. Since $B_3$ is rationally fibered we can define two sections $\mathcal{S}_-$ and $\mathcal{S}_+$ of $B_3$. These two sections are related via 
\begin{equation}\label{sectionrelations}
 \mathcal{S}_+ = \mathcal{S}_- + \rho^*c_1(\mathcal{L}) \,,
\end{equation} 
where $\mathcal{L}$ is the line bundle describing twist of the $\mathbb{P}^1$-fiber of $B_3$ over $B_2$. This twist bundle is closely related to the choice of the heterotic gauge bundle, and for the simple case of standard embedding, one needs to choose $c_1(\mathcal{L}) = 6c_1(B_2)$~\cite{Friedman:1997ih}. In this case, an $E_8$ gauge theory in F-theory is realized on $\mathcal{S}_-$ whereas the visible $E_6$ gauge theory is realized on $\mathcal{S}_+$. The respective gauge couplings are then given by 
\begin{equation}\label{gaugecouplingF}
 \frac{1}{g_{E_8}^2} = 2\pi \mathcal{V}_{\mathcal{S_-}}\,,\qquad \frac{1}{g_{E_6}^2} = 2\pi \mathcal{V}_{\mathcal{S}_+}\,. 
\end{equation} 
Since $\mathcal{S}_-$ is a copy of $B_2$ it is clear from \eqref{sectionrelations} that, in the adiabatic limit for the fibration $\rho$, the gauge couplings of both gauge groups approximately agree. This limit is, however, nothing but the weak coupling limit for the heterotic string in which, indeed, the gauge couplings of the gauge groups in both $E_8$ factors are given by the tree level expression in \eqref{treelevelgaugekinetic} and thus determined by the heterotic 4d dilaton. Moving away from the strict adiabatic limit, the two gauge couplings differ by the volume of the vertical divisor 
\begin{equation}
  \frac{1}{g_{E_6}^2}- \frac{1}{g_{E_8}^2} = 2\pi \mathcal{V}_{\rho^* c_1(\mathcal{L})} \,,
\end{equation} 
which, as demonstrated in \cite{Klaewer:2020lfg}, is the F-theory avatar of the non-holomorphic threshold corrections to the heterotic gauge coupling. Therefore, the classical F-theory geometry captures a subset of the perturbative corrections to the heterotic gauge coupling. In the dual F-theory frame, the heterotic string coupling is given by a linear combination of divisor volumes
\begin{equation}\label{eq:ghetF}
 \frac{1}{g_{\rm het}^2} = \mathcal{V}_{\mathcal{S}_-} + a\, \mathcal{V}_{\rho^* c_1(\mathcal{L})}\,,
\end{equation} 
for some $a>0$. Given this identification, we can now discuss the strong coupling singularity for the heterotic gauge group from the F-theory perspective. Therefore, recall that, for the perturbative heterotic string compactified on $X_3$, the strong coupling singularity is encountered if $g_{\rm het}$ is kept constant while the K\"ahler moduli of $X_3$ are scaled to infinity (in heterotic string units). In particular, this includes the volumes of the curves in the base $B_2$, which, via duality to F-theory, translate into the volumes of $\rho$-vertical divisors. In terms of \eqref{eq:ghetF} the heterotic limits thus correspond to taking $\mathcal{V}_{\rho^* c_1(\mathcal{L})}\rightarrow \infty$ while keeping the LHS constant. This is, however, only possible if $\mathcal{V}_{\mathcal{S}_-}$ decreases at the same time such that $\mathcal{V}_{\mathcal{S}_-}=0$ is reached after traversing a finite distance in the field space. Via \eqref{gaugecouplingF}, this corresponds to the strong coupling singularity for the hidden $E_8$ gauge group. 

From this dual perspective, the strong coupling singularity thus corresponds to the classical boundary of the K\"ahler cone of $B_3$ along which a rigid, effective divisor, in this case $\mathcal{S}_-$, shrinks to zero size.  As mentioned above, this classical F-theory perspective accounts for both the heterotic tree-level and threshold effects. As in the perturbative heterotic analysis, the classical F-theory analysis suggests the existence of a $\text{codim}_{\mathbb{R}}=1$ singularity corresponding to the boundary of the K\"ahler cone along which the divisor $\mathcal{S}_-$ shrinks to zero size. 

Let us stress that \eqref{eq:ghetF} only captures the effect of the heterotic K\"ahler moduli of the base $B_2$ on the heterotic couplings. The heterotic threshold corrections discussed in the previous sections, on the other hand, treat all K\"ahler moduli equally which includes the modulus of the elliptic fiber of $X_3$. As reviewed above, the volume of the elliptic fiber of $X_3$ on the heterotic theory gets mapped to a complex structure modulus of $X_4$ on the F-theory side of the duality. Therefore, from the perspective of the F-theory compactification, the heterotic threshold corrections predict a mixing between the K\"ahler moduli of $B_3$ and the complex structure moduli of $X_4$. Such mixing is absent at the classical level but can be present at the quantum level as analyzed in \cite{Wiesner:2022qys}. In particular, in that reference, it was shown that the mixing of the complex structure and K\"ahler moduli indeed leads to a strong coupling singularity for the gauge theory realized on $\mathcal{S}_-$ -- at least in the case of heterotic standard embedding. The strong coupling singularity that we obtained from the perturbative heterotic analysis in the previous section is hence related via duality to the strong coupling singularity discussed in \cite{Wiesner:2022qys}.\\

The situation in F-theory is reminiscent of strong coupling singularities in Type IIA compactifications on Calabi--Yau threefolds. In this setup, the vector multiplet moduli space of the resulting $\cN=2$ theory in four dimensions is, at the classical level, identical to the K\"ahler moduli space of $X_3$. In this classical moduli space, sub-manifolds of $X_3$ shrink to zero size along special loci. If the sub-manifold is a divisor shrinking to a point, the respective locus in moduli space corresponds classically to a boundary of the K\"ahler moduli space. Furthermore, along this boundary, the $U(1)$ gauge theory obtained by reducing the Type IIA 3-form over the curve dual to the shrinking divisor becomes strongly coupled. If instead of Type IIA, we consider M-theory compactified on the same $X_3$, this boundary corresponds to a 5d SCFT locus in moduli space. On this locus, a string obtained by an M5-brane wrapping the shrinking divisor becomes tensionless together with a tower of BPS states obtained by wrapping M2-branes on curves inside the shrinking divisor that become massless. Since in M-theory on Calabi--Yau threefolds, the vector multiplet moduli space is real and does not receive any quantum corrections, the SCFT locus is indeed a $\text{codim}_{\mathbb{R}}=1$ boundary of the moduli space. 
However, in the case of a Type IIA compactification on the same manifold, the vector multiplet moduli space gets complexified. It hence cannot have a $\text{codim}_{\mathbb{R}}=1$ boundary. Instead, in Type IIA Calabi--Yau compactifications, worldsheet instantons correct the moduli space geometry and partially resolve the strong coupling singularity as can be seen explicitly using mirror symmetry. To see the physical effect of this resolution, one can consider the central charges of BPS particles in the effective $\cN=2$ obtained by wrapping a D4-brane on the shrinking divisor and D2-branes wrapping curves inside this divisor. These central charges can be computed via mirror symmetry with the result that the central charge of the D4-brane wrapping the shrinking divisor vanishes at a $\text{codim}_{\mathbb{C}}=1$ conifold singularity that can be identified with the residual K\"ahler cone boundary. On the other hand, the central charges of all D2-branes wrapping curves inside this divisor remain finite due to the effects of the worldsheet instantons. Thus, there remains a strong coupling singularity in the vector multiplet moduli space of Type IIA Calabi--Yau compactifications at $\text{codim}_{\mathbb{C}}=1$. However, due to the non-perturbative effects of the worldsheet instantons, the singularity is much less severe than its $\text{codim}_{\mathbb{R}}=1$ analog in five-dimensional compactifications of M-theory. In particular, the moduli space of Type IIA on $X_3$ does not end at the strong coupling singularity, but there is merely a phase transition into a non-geometric or orbifold phase. \\

The classical boundary of the K\"ahler moduli space of Calabi--Yau threefold compactifications of Type IIA string theory can be viewed as the analog of the classical strong coupling singularity for the gauge theory on the section $\mathcal{S}_-$ in the $\cN=1$ compactification of F-theory. Based on the insights from the $\cN=2$ vector multiplet moduli space, we also expect that non-perturbative corrections to the $\cN=1$ field space resolve the strong coupling singularity discussed above. Computing these effects directly in the perturbative heterotic or F-theory duality frame is currently out of reach: The relevant corrections arise from non-BPS instanton corrections to the K\"ahler potential in F-theory which as of now cannot be calculated reliably due to the lack of powerful techniques. In addition, there are perturbative corrections to K\"ahler potential which compete with the non-perturbative effects in the regime of interest to us. To make any concrete statements, we need to describe the problem in which both the perturbative and non-perturbative effects can be computed explicitly. This is, indeed, possible if we consider the strong-coupling description of the heterotic string in terms of M-theory on an interval. Using this duality frame, we first recover the perturbative strong coupling singularity in the next section and subsequently investigate the effect of non-perturbative corrections in section~\ref{sec:nonpert}.

\section{Classical Domain Wall Solution}\label{sec:classDW}
In the previous section, we considered threshold corrections to the gauge coupling in compactifications of the heterotic $E_8\times E_8$ string on Calabi--Yau threefolds and their avatars in dual F-theory compactifications. To understand the fate of the strong coupling singularity for one of the $E_8$ factors from a non-perturbative perspective our strategy is to utilize a dual strong-coupling description of the heterotic string for which non-perturbative effects can be computed. It is well-known that the strong coupling limit of the ten-dimensional heterotic string is M-theory on $S^1/\mathbb{Z}_2$. Compactifying both sides of the duality on a Calabi--Yau threefold identifies the heterotic string on $X_3$ with M-theory on $X_3\times S^1/\mathbb{Z}_2$. As a first step, we now demonstrate how the strong coupling singularity of the heterotic string is reproduced in this latter duality frame.

As in the previous section, we exclusively focus on the K\"ahler sector of $X_3$. Compactifying M-theory on this threefold gives a five-dimensional theory with $\cN=1$ supersymmetry. The massless spectrum of this theory is given by $h^{2,1}+1$ hypermultiplets and $h^{1,1}(X_3)$ vector multiplets. In the following, we mainly focus on the universal hypermultiplet. Alternatively, we can consider the Calabi--Yau threefold $X_3$ to be rigid, i.e., $h^{2,1}=0$. The moduli space of this universal hypermultiplet is classically given by a quaternionic manifold with coset structure $\cM_Q = SU(2,1)/SU(2) \times U(1)$ which is spanned by four real coordinates $q^u=(V,\sigma, \zeta, \tilde{\zeta})$. These can, respectively, be identified as the volume of $X_3$, the axion dual to the M-theory three-form in five dimensions, and the periods of $C_3$ over the fundamental $A$- and $B$-three-cycles in $X_3$. Out of these fields, we define the following combinations 
\begin{equation}\label{eq:Sandxi}
S= V+ \frac{i}{4}\left(\sigma + \xi  \tilde\xi \right)\,,\qquad \xi= \zeta\,,\qquad \tilde{\xi} = \tilde{\zeta} - \tau \zeta \,. 
\end{equation}
Here $\tau$ is the ratio of the two periods of the rigid Calabi--Yau appearing in the classical, holomorphic prepotential of the rigid threefold, 
\begin{equation}
 F(X^0) = -\frac{\tau}{2} (X^0)^2\,. 
\end{equation} 
In terms of these fields the classical K\"ahler potential for the complex field $S$ is given by 
\begin{equation}\label{eq:classicalKQ}
K = -\log (S+\bar{S})\,. 
\end{equation} 
The resulting five-dimensional action can then be written as 
\begin{equation}\label{eq:S5d}
S_{\rm 5d} = S_{\rm hyper} + S_{\rm grav} \,,
\end{equation}
with 
\begin{equation}\label{Shyper}
 S_{\rm hyper} = -\frac{M_{\rm pl,5}^3}{2} \int \sqrt{-g} \left[h_{uv}\nabla_\alpha q^u \nabla^\alpha q^v \right]\,,
\end{equation} 
and 
\begin{equation}
 S_{\rm grav} = -\frac{M_{\rm pl,5}^3}{2} \int\sqrt{-g} \left[R + \frac{3}{2} \cF_{\alpha \beta} \cF^{\alpha \beta} + \frac{1}{\sqrt{2}} \epsilon^{\alpha \beta \gamma \delta\epsilon} \cA_\alpha \cF_{\beta \gamma} \cF_{\delta \epsilon}\right]\,.\\
\end{equation}
Here $M_{\rm pl,5}$ is the five-dimensional Planck mass, $h_{uv}$ is the metric on the moduli space spanned by the universal hypermultiplet as derived from the K\"ahler potential \eqref{eq:classicalKQ}, and $\cF$ is the field strength of the gauge fields $\cA$ in the 5d bulk vector multiplets.  Compactifying this five-dimensional theory on an additional interval $S^1/\mathbb{Z}_2$ yields an effective four-dimensional theory with $\cN=1$ supersymmetry. The fixed loci of the $\mathbb{Z}_2$ orbifold action correspond to space-time filling 9-branes that wrap the entire $X_3$ on which the perturbative heterotic gauge groups in the two $E_8$ factors are realized. 

The presence of the orbifold fixed planes gives an additional boundary contribution to the effective action that originates from the 10-dimensional Yang--Mills action
\begin{equation}
\begin{aligned}
 S_{{\rm YM}, (10)} = -\frac{M_{\rm pl, 11}^6}{(32\pi)^{1/3}} & \left[\int_{\cM_{10}^{(1)}} \sqrt{-g_{10}}\left(\tr(F^{(1)})^2 - \frac12 \tr R^2 \right)\right. \\
 & + \left. \int_{ \cM_{10}^{(2)}} \sqrt{-g_{10}}\left(\tr(F^{(2)})^2 - \frac12 \tr R^2 \right)\right]\,,
\end{aligned}
\end{equation}
where $F^{(a)}$, $a=1,2$, are the field strength of the gauge theories realized on the 9-branes located at $\cM_{10}^{(a)}$ and $M_{\rm pl,11}$ is the 11-dimensional Planck scale. Due to the presence of these 9-branes, the Bianchi identify for the $G_4$ form in 10d is modified to be 
\begin{equation}
 {\rm d}G_4 = -\frac{1}{2\sqrt{2}\pi} \left(\frac{1}{4\pi M_{\rm pl,11}^{9}}\right)^{2/3} \left[J^{(1)} \delta(y) + J^{(2)} \delta(y-2 \pi \rho) \right]\,,
\end{equation} 
where $y$ is the coordinate of the interval $S^1/\mathbb{Z}_2$ of length\footnote{The length of the interval is typically given by $\pi \rho$. Here we add the extra factor of 2 for later convenience such that prior to orbifolding the M-theory circle has radius $2\rho$ in M-theory units.}  $\ell_{S^1/\mathbb{Z}_2}M_{\rm pl, 11}= 2\pi \rho$ and the currents $J^{(a)}$ are given by 
\begin{equation}\label{Ji}
 J^{(a)} = \left(\tr F^{(a)} \wedge F^{(a)} -\frac12 \tr R\wedge R \right)\,. 
\end{equation} 
Let us now compactify this theory on a Calabi--Yau threefold. Then, a non-zero second Chern class of the threefold induces a non-trivial source $J^{(a)}$ unless a non-trivial gauge bundle on the orbifold planes cancels it. In this work, we are interested in domain wall solutions that are dual to perturbative heterotic string theories without NS5-brane, implying that RHS of the Bianchi identity \eqref{het:Bianchi} vanishes. More specifically, we want to focus on the case of heterotic standard embedding, in which case the Bianchi identity is satisfied for 
\begin{equation}\label{standardembedding}
\tr F^{(1)}\wedge \tr F^{(1)} = \tr R\wedge R\,,
\end{equation}
and the trivial gauge bundle in the second $E_8$ factor. However, this implies that the sources $J^{(a)}$ in the two orbifold planes are non-zero. In fact, they are the same with opposite signs. We can interpret this as an M5-brane charge localized on $\cM_{10}^{(a)}$:
\begin{equation}
 Q_{M5}^{(i)}= \frac{(-1)^{a+1}}{2}  \tr R\wedge R\,. 
\end{equation} 
The M5-brane charge induces a non-trivial potential for the scalar fields given by 
\begin{equation}\label{scalarpotential}
\mathcal{V} = -\frac{1}{3} e^{2K} \alpha^2 \,,
\end{equation}
where the parameter $\alpha$ depends on the $h^{1,1}$ vector multiplets 
\begin{equation}\label{alpha}
 \alpha = -\frac{\alpha_0}{V^{1/3}} \int_{X}  \omega \wedge c_2(X_3)\,,
\end{equation} 
where $\omega$ the K\"ahler form on $X_3$ and $\alpha_0$ is a constant. Importantly, $\alpha$ is independent of an overall rescaling of the K\"ahler form such that it does not depend on the universal hypermultiplet. The dependence of the scalar potential on the universal hypermultiplet is hence entirely encoded in the prefactor containing the K\"ahler potential for the universal hypermultiplet. Denoting the loci fixed by the orbifold action by $\cM_4^{(a)}$ for $a=1,2$, the boundary action after dimensional reduction is given by 
\begin{equation}
\begin{aligned}
 S_{\rm bound} =&  -\frac{M_{\rm pl,5}^3}{2} \left(-2\sqrt{2} \int_{\cM_4^{(1)}} \sqrt{-g} e^{K} \alpha+2\sqrt{2} \int_{\cM_4^{(2)}} \sqrt{-g} e^{K} \alpha  \right) \\ &- \frac{1}{16\pi \alpha_{\rm GUT}} \sum_{a=1}^2 \int_{\cM_4^{(a)}} \sqrt{g} V \tr F_{\rm \mu \nu}^{(a)^2}\,,
 \end{aligned}
\end{equation} 
where $\alpha_{\rm GUT}$ is the bare coupling of the gauge theory on the domain walls.
Due to the presence of the M5-brane charge, i.e., the non-trivial $\alpha\neq 0$, the 5d configuration with constant field profile along $S^1/\mathbb{Z}_2$ is not a solution to the equations of motion. BPS solutions to the classical equations of motion have been discussed in detail in the literature, see, e.g., \cite{Lukas:1998yy}, for which we review the main aspects in the following. To that end, we first re-express the scalar potential $\mathcal{V}(\phi)$ in terms of a superpotential $W$. In $d$ dimensions these are related as \cite{Skenderis:1999mm} via 
\begin{equation}
 \cV= 2(d-2) \left[(d-2) |\nabla W|^2 -(d-1) |W|^2\right]\,. 
\end{equation} 
The scalar potential in \eqref{scalarpotential} can be reproduced with the superpotential 
\begin{equation}
 W = - \frac{1}{3} e^{K} \alpha\,. 
\end{equation}
For the domain wall solution we use the following ansatz for the 5d metric
\begin{equation}\label{metricansatz}
 {\rm d}s_5^2 = e^{2a} {\rm d}x^\mu {\rm d}x^\nu \eta_{\mu \nu} + e^{8a} {\rm d}y^2 \,,\qquad V=V(y)\,,
 \end{equation}
 where $y$ is the coordinate along the interval. The BPS equations then take the form \cite{Cvetic:1992bf,Cvetic:1994ya,Cvetic:1996vr,Skenderis:1999mm} 
 \begin{equation}\label{BPSequations}
 \begin{aligned}
 \partial_y a(y) &= \mp \frac{1}{4} e^{4a} W(S,{\bar S}) \,,\\ 
 \partial_y S(y) & = \pm \frac{3}{4} e^{4a} K^{S \bar{S}} \partial_{\bar S} W(S,\bar{S})\,,
 \end{aligned}
 \end{equation}
 where the complex coordinate $S$ is related to the volume $V$ of $X_3$ as in \eqref{eq:classicalKQ}. A solution to these equations is given by 
 \begin{equation}\label{classicalsolution}
  a_0 e^{6a}  = \text{Re}\,S(y)= V(y)\,, \quad V(y) = V_0 H(y)^3\,,
 \end{equation}
 for some integration constants $V_0$ and $a_0$ and a harmonic function $H(y)$ satisfying 
 \begin{equation}
  \partial_y^2 H(y) = \frac{2\sqrt{2}}{3}\alpha \left(\delta(y) - \delta(y-2 \pi \rho) \right)\,,
 \end{equation} 
to account for the non-trivial sources $J^{(a)}$ localized on the orbifold fixed planes as defined in \eqref{Ji}. A solution to this differential equation is given by 
 \begin{equation}\label{Hofy}
 H(y)= \frac{\sqrt{2}}{3}\alpha |y| +c_0\,,
 \end{equation} 
 with $c_0$ a constant. The constants $V_0$ and $c_0$ determine the gauge coupling of the $E_6$ gauge theory realized on the visible brane located at $y=0$ as 
 \begin{equation}
 \frac{1}{g_{E_6}^2} = V(0)= V_0 c_0^3 \,, 
 \end{equation} 
 whereas the gauge coupling on the hidden $E_8$ domain wall is given by 
 \begin{equation}
 \frac{1}{g_{E_8}^2} = V(2\pi \rho) \,. 
 \end{equation}
 Importantly, the harmonic function $H(y)$ vanishes at 
 \begin{equation}
  y_{\rm max} = -\frac{3c_0}{\sqrt{2} \alpha} \,. 
 \end{equation} 
such that 
 \begin{equation}
V(y_{\rm max})= a_0 e^{6a}|_{y=y_{\rm max}}= 0\,.
 \end{equation} 
This implies that if the position of the hidden domain wall is chosen to be $2 \pi \rho = y_{\rm max}$, the gauge coupling and the warp factor become singular at this domain wall. For this choice of the brane position, we thus encounter a strongly coupled end-of-the-world brane.

This is, however, just the classical picture. In particular, the solution \eqref{classicalsolution} to the BPS equations \eqref{BPSequations} crucially relies on the classical expression for the K\"ahler potential $K$ in \eqref{eq:classicalKQ}. Quantum corrections to the classical moduli space geometry of the universal hypermultiplet, on the other hand, become important away from the $V\gg1$ regime and, therefore, may significantly alter the solution if $2\pi \rho \sim y_{\rm max}$. The next section will discuss the corrections to the domain wall solution. For the moment, we stick to the classical solution and compare the result with the perturbative heterotic analysis of the previous section. 

\subsection{Heterotic/M-theory duality map}
We first provide a qualitative map between the moduli of the heterotic string on $X_3$ and the moduli of the domain wall solution for M-theory on $X_3\times S^1/\mathbb{Z}_2$. In particular, we focus only on the moduli that are involved in the running of the gauge coupling. On the heterotic side, these are the (complexified) string coupling $S_{\rm het}$ and the overall (complexified) K\"ahler modulus $T$ of $X_3$. On the M-theory side, the moduli are the value of the dilaton field at the position of the visible domain wall, $S_0=S(0)$, and the length of the interval $\ell_{S^1/\mathbb{Z}_2} M_{\rm pl,11} = 2 \pi \rho$. In the four-dimensional theory, this latter modulus pairs up with the Wilson line of the five-dimensional graviphoton on $S^1/\mathbb{Z}_2$ to give a complex scalar field. To match the fields of the M-theory domain wall with the heterotic moduli, we first recall that on the heterotic side, we have the identity 
\begin{equation}
 \frac{M_{\rm pl,4}^2}{M_h^2} = \frac{\kappa}{g_s^2} (\text{Re}\,T)^3 = \text{Re}\,S_{\rm het}\,,
\end{equation}
where the constant $\kappa$ is related to the triple intersection numbers of $X_3$ and $g_s$ is the string coupling of the ten-dimensional heterotic string. Upon heterotic/M-theory duality, the string coupling is related to the length of the Horawa--Witten interval as 
\begin{equation}
 g_s = \rho^{3/2} \,,
\end{equation} 
and the four-dimensional Planck scale in M-theory units is given by 
\begin{equation}
  \frac{M_{\rm pl,4}^2}{M_{\rm pl,11}^2} = 2\pi \,\text{Re}\,S_0 \,g_s^{2/3}\,.
\end{equation}
We can thus identify
\begin{equation}\label{identification}
 2\pi \text{Re}\,S_0= \text{Re}\,S_{\rm het} \,,\qquad \rho = \frac{1}{(\text{Re}\,S_{\rm het})^{1/3}} \text{Re}\,T\,,
\end{equation} 
where we used $M_h/M_{\rm pl,11} = g_s^{1/3}$.  Notice that, in total, the domain wall solution in \eqref{classicalsolution} has three free parameters relevant for our discussion: $V_0, c_0$ and $\rho$. The above identification determines $\rho$ in terms of the heterotic variables. On the other hand $\text{Re}\,S_0$ is given by 
\begin{equation}\label{identificationS0}
 \text{Re}\,S_0 = c_0^3 V_0 \,, 
\end{equation}
which, via \eqref{identification}, relates the combination of $c_0^3 V_0$ to $\text{Re}\,S_{\rm het}$. This leaves us with one additional parameter of the domain wall solution that is not yet fixed in terms of the heterotic variables. From the heterotic perspective, we have so far only considered the string coupling and the geometric moduli of $X_3$. In addition, the perturbative heterotic string has moduli associated with the non-trivial gauge bundles in the two $E_8$ factors. Recall that, in this work, we focus on the case of standard embedding for which we choose the bundle $V_a$ in the $a$-th $E_8$ factor to satisfy 
\begin{equation}
c_2(V_1) = c_2(X_3) \,,\qquad c_2(V_2)=0\,. 
\end{equation} 
This constraint does not uniquely determine $V_1$. Instead, there is a non-trivial moduli space for $V_1$, which only reduces to the tangent bundle of $X_3$ \eqref{2,2locus} along the $(2,2)$ locus in the bundle moduli space for which the worldsheet theory of the perturbative heterotic string has enhanced $(2,2)$ supersymmetry. Deformations within this moduli bundle moduli space break the worldsheet supersymmetry to $(0,2)$, affecting the computation of the threshold corrections. Accordingly, the discussion of the threshold correction in section~\ref{sec:perthet} is only valid on this $(2,2)$ locus. As shown in the following, the map between the heterotic threshold corrections computed on the $(2,2)$ locus and the profile $S(y)$ for the domain solution fixes the remaining combination of $c_0$ and $V_0$. This combination hence determines the location of the $(2,2)$ locus in the heterotic bundle moduli space in terms of the moduli of the domain wall solution. 

Using the identification \eqref{identification}, we can compare the running of the heterotic gauge coupling induced by the perturbative threshold corrections to the running of the dilaton in the Horava--Witten domain wall solution. Focusing on the heterotic standard embedding, the gauge coupling of the unbroken $E_8$ gauge factor is given by  
\begin{equation}
\frac{1}{g_{E_8}^2}= \text{Re}\,S_{\rm het}  + \frac{1}{16\pi^3} \Delta_{a}(T, \bar{T}) + \dots\,. 
\end{equation} 
Here, the threshold correction is given in \eqref{def:Deltaa}, which for $T\rightarrow \infty$ has an expansion as in \eqref{leadingcorrections}. Recall from \eqref{classicalsolution} and \eqref{Hofy} that, on the M-theory side, the profile for $V$ as a function of $y$ is given by 
\begin{equation}\label{Vofy} 
 V(y) = V_0\left(\frac{\sqrt{2}}{3} \alpha |y| +c_0\right)^3 \,,
\end{equation}
where $V_0$ and $c_0$ are constants. To match the running of the dilaton in the domain wall solution to the perturbative threshold corrections, we expand $V(2\pi \rho)$ for large $c_0$ as
\begin{equation} 
 V(2 \pi \rho) = V_0 c_0^3 +2 \pi  \rho \sqrt{2} V_0 c_0^2 \alpha + \mathcal{O}(c_0)\,. 
\end{equation} 
Via \eqref{identificationS0}, the $c_0\gg 1$ limit corresponds to the perturbative limit for the heterotic string. Using \eqref{identification} we trade $\rho$ for $\text{Re}\,T$ to obtain
\begin{equation}
 V(\rho) = \frac{1}{2\pi} \text{Re}\,S_{\rm het} +( 2 \pi )^{2/3}\sqrt{2} V_0^{2/3} c_0 \alpha \kappa^{1/3} \text{Re}\,T  + \dots \,. 
\end{equation} 
The comparison with the leading corrections to the gauge coupling of the hidden $E_8$ on the $(2,2)$ locus identifies the combination $V_0^{2/3} c_0$ with
\begin{equation}
 V_0^{2/3} c_0 = -\frac{b_a \gamma}{\sqrt{2} (2\pi)^{4/3} \kappa^{1/3}}\,,
\end{equation} 
where the extra minus sign comes from the definition of $\alpha$ in terms of the second Chern class in \eqref{alpha}. In other words, the two parameters $V_0$, $c_0$ can be viewed as encoding both the bare coupling as well as the $\beta$-function coefficient of the heterotic gauge theory in the case of heterotic standard embedding on the (2,2) locus.  Notice that this identification was possible due to the common appearance of the integrated second Chern class in both the heterotic threshold corrections and the running of the dilaton in the domain wall solution. In fact, the domain wall solution provides evidence that also for general Calabi--Yau threefolds the leading threshold correction in the $\text{Re}\,T\gg1$ limit is controlled by the second Chern class of the Calabi--Yau. 

We stress that the domain wall on the M-theory side is a purely classical solution to the equation of motion. Still, it captures the one-loop effects of the heterotic string as it reproduces the heterotic threshold corrections to the gauge coupling. This is similar to the observation in \cite{Klaewer:2020lfg} that the classical F-theory geometry also captures certain threshold corrections to the heterotic gauge coupling, as reviewed in the previous section.

\subsection{A Comment on the Six-dimensional Case}\label{sec:6dcomment}
Our analysis so far could have been similarly carried out in six dimensions if we considered the heterotic string on K3 or its dual given by M-theory on $K3\times S^1/\mathbb{Z}_2$. Also, in this setup, we can consider the standard embedding for the gauge bundles corresponding to the instanton embedding $(24,0)$ in the two $E_8$ gauge groups of the heterotic string. On the M-theory side of the duality, this similarly induces a non-zero 5-brane charge in the 9-branes located at the orbifold fixed points at the end of the M-theory interval. Thus, in this case, we would similarly get a running solution for which the volume of the K3 changes as a function of the coordinate on the interval. Just as in the four-dimensional case, the unbroken $E_8$ gauge theory in the 6d heterotic theory receives threshold corrections 
\begin{equation}
 \frac{1}{g_{\rm E_8}^2 M_{\rm het}^2} = S_{\rm het,6d} - \frac{b}{16\pi^3} \int_{K3} c_2 \, +\dots \,,
\end{equation}
for some constant $b$. Notice, however, that unlike in the 4d $\cN=1$ compactification of the heterotic string, the threshold corrections for the heterotic string on $K3$ are independent of all the geometric moduli of $K3$. This is, in fact, not surprising since the moduli space of the 6d $\cN=(1,0)$ theory factorizes into a tensor- and the hypermultiplet sector:
\begin{equation} 
\cM_{\cN=(1,0)} = \cM_{\rm tensor} \times \cM_{\rm hyper}\,. 
\end{equation} 
For the perturbative heterotic string on K3, the geometric moduli of K3 and the bundle moduli are part of the hypermultiplet sector, whereas the heterotic dilaton is the single tensor of the theory. Due to the factorization of the moduli space threshold, corrections to the gauge coupling of the perturbative heterotic gauge group must be moduli-independent. Still, in the M-theory picture, we encounter a strong coupling singularity for the hidden 9-brane if we increase the length of the interval to a critical value. As in the previous discussion, this corresponds to an end-of-the-world brane since the warp factor of the domain wall also diverges at the location of the 9-brane as can be seen by explicitly solving the 6d BPS domain wall equations.

Unlike in the four-dimensional setups, which are the main focus of this paper, the strong coupling singularity in the six-dimensional case is well understood. Again, it is helpful to consider the F-theory dual. We, therefore, recall the following duality for the heterotic string on K3~\cite{Morrison:1996na, Morrison:1996pp}
\begin{equation}\label{hetFdual}
\text{Het. instanton embedding}\; (12-n,12+n) \qquad \longleftrightarrow \qquad \text{F-theory on}\;T^2 \hookrightarrow X_3 \rightarrow \mathbb{F}_n\,. 
\end{equation}
The strong coupling singularity for the perturbative heterotic gauge theory with standard $(24,0)$-embedding corresponds to the locus in the K\"ahler moduli space where the base $\mathbb{P}^1_b$ of $\mathbb{P}^1_f\hookrightarrow \mathbb{F}_{12} \rightarrow \mathbb{P}^1_b$ vanishes. This corresponds to a real co-dimension one boundary of the K\"ahler moduli space of $\mathbb{F}_{12}$ which can be identified with the tensor branch of the 6d $\cN=(1,0)$ theory. Unlike the chiral field space of the 4d $\cN=1$ theory, the tensor branch is real and hence can have a real co-dimension one boundary. Therefore, the strong coupling singularity in 6d does not need to be resolved. Due to the absence of non-perturbative corrections to the tensor branch, the strong coupling singularity in the 6d theory cannot be resolved such that, indeed, it corresponds to a boundary of the tensor branch. In particular, the tensor branch cannot be extended beyond the strong coupling singularity for the hidden gauge group.\footnote{In the six-dimensional case, it is also clear what happens for other instanton embeddings. The other extreme case is the symmetric $(12,12)$-embedding. From \eqref{Ji}, we infer that, in this case, there are no 5-brane sources localized in the Horava--Witten 9-branes, and no non-trivial profile for the dilaton is induced. By \eqref{hetFdual}, the F-theory dual of this instanton embedding is F-theory with base $B_2=\mathbb{F}_0=\mathbb{P}^1\times \mathbb{P}^1$. The absence of a non-trivial profile for the dilaton in the Horava--Witten setup implies the absence of the strong coupling singularity for symmetric embedding. From the dual F-theory perspective, this is manifest in the absence of a finite distance boundary of the tensor branch of F-theory on $\mathbb{F}_0$: Due to the trivial fibration, all boundaries of this tensor branch are weak-coupling limits for heterotic strings such that, indeed, there is also no strong coupling singularity in F-theory as is consistent with the M-theory picture.}

The difference between the six- and four-dimensional setups can also be understood from the perspective of M-theory on $K3\times S^1/\mathbb{Z}_2$. In this case, the real co-dimension one strong coupling singularity manifests itself as the end-of-the-world brane at which the warp factor diverges. The moduli space of 7d parent theory corresponding to M-theory on K3 is given by 
\begin{equation}
 \cM_{7d} = O(\Gamma_{3,19})\Big\backslash \frac{O(3,19)}{O(3)\times O(19)} \times \mathbb{R}_+\,.
\end{equation}
The first factor corresponds to the geometric moduli of the K3, and the second factor to the overall volume of K3. After compactifying on $S^1/\mathbb{Z}_2$ further down to 6d, the induced 5-brane charge induces a profile only for the overall volume of K3. Since the moduli space for this modulus is exactly $\mathbb{R}_+$, there are no corrections to be taken into account, and the classical domain wall solution is exact. Therefore, the end-of-the-world brane with diverging warp factor also persists at the quantum level in 6d. This is consistent with the fact that we cannot extend the 6d moduli space beyond the strong coupling singularity. As we will see in the following section, the situation is markedly different in four dimensions. 

\section{Instanton Corrections to Domain Wall}\label{sec:nonpert}
So far we discussed the classical solution to the domain wall equations of motion and matched it with the perturbative corrections to the heterotic gauge coupling. As mentioned before, the classical domain wall solution gets corrected at the quantum level due to corrections to the metric for the universal hypermultiplet that is running along the direction of the interval. These corrections are already present in the five-dimensional $\cN=1$ parent theory and can be split into perturbative and non-perturbative corrections. Notably, the hypermultiplet moduli space of M-theory on a Calabi--Yau threefold is the same as the hypermultiplet moduli space of Type IIA string theory compactified on the same threefold. In the Type IIA description, perturbative corrections to the universal hypermultiplet moduli space can only arise at one-loop which can be computed explicitly and are given in terms of the Euler characteristic of the Calabi--Yau threefolds~\cite{RoblesLlana:2006ez}. More interesting are the non-perturbative corrections which originate from Euclidean 5-brane instantons wrapping the entire Calabi--Yau threefold and Euclidean 2-brane instantons wrapping 3-cycles. Since we are focusing exclusively on the universal hypermultiplet, the effect of the 5-brane instantons is particularly relevant. In Type IIA string theory, the effect of NS5-brane instantons on the hypermultiplet moduli space metric has, at least partially, been computed in \cite{Alexandrov:2006hx,Alexandrov:2009vj,Alexandrov:2010ca,Alexandrov:2014mfa,Alexandrov:2014rca,Alexandrov:2023hiv}. The computations to obtain the NS5-brane instanton corrections to the hypermultiplet moduli space of Type IIA Calabi--Yau threefold compactifications are somewhat involved. We will, therefore, not review all the details here but give an overview of their effect, which suffices to understand how 5-brane instantons affect the domain wall solution discussed in the previous section. 

Following this approach, the \emph{full} set of corrections to the domain wall solution of the 5d $\cN=1$ effective theory governed by the two-derivative action \eqref{eq:S5d} can, in principle, be computed. After dimensional reduction, this yields a 4d $\cN=1$ theory dual to the heterotic/F-theory setups discussed in section~\ref{sec:perthet}. As discussed at the end of that section, computing such effects directly in the 4d $\cN=1$ theory is notoriously difficult since the non-perturbative effects arise from non-BPS instantons and higher-order perturbative corrections might be equally important. Using the domain wall perspective, we calculate corrections to the 4d $\cN=1$ theory in terms of known corrections to the 5d $\cN=1$ hypermultiplet sector. Crucially, in the 5d $\cN=1$ hypermultiplet sector, the instanton corrections are BPS, and higher-order perturbative corrections are absent due to the enhanced supersymmetry of the parent theory, thus making these quantum effects amenable for explicit computations. 

In the following, we again focus on the universal hypermultiplet. The scalar fields in the universal hypermultiplet are defined as in \eqref{eq:Sandxi}. The identification in \eqref{eq:Sandxi} is valid at the classical level, but the relation between the (complex) fields $(S, \xi, \tilde{\xi})$ and the (real) fields $(V,\sigma, \zeta, \tilde{\zeta})$ receives correction at the quantum level. In section~\ref{sec:corrections}, we first review the computations of the NS5-brane/M5-brane instantons to the hypermultiplet moduli space and apply these results to the domain wall solution in section~\ref{sec:corrDW}.

\subsection{Instanton Corrections to Hypermultiplet Moduli Space}\label{sec:corrections}
To introduce the necessary background on corrections to the hypermultiplet moduli space $\cM$, we follow \cite{Alexandrov:2008gh,Alexandrov:2009vj,Alexandrov:2011va,Alexandrov:2023hiv}.  Let us stress that the hypermultiplet moduli space $\cM$ is a quaternionic K\"ahler manifold, and a description in terms of a K\"ahler potential does not exist. Still, there is a way to connect the metric on $\cM$ to a K\"ahler metric on a different space, called the twistor space. This twistor space $\pi:\CZ \rightarrow \cM$  can be viewed as a $\mathbb{P}^1$ bundle over $\cM$. The metric on $\CZ$ is related to the metric on $\cM$ as 
\begin{equation}\label{eq:metrictwistor}
{\rm d}s^2_{\CZ} = \frac{|Dt|^2}{(1+t\bar{t})^2} +\frac{\Lambda}{12} {\rm d} s^2_{\cM} \,. 
\end{equation}
Here $\Lambda$ is the constant, negative curvature of the hypermultiplet moduli space, $t$ is the stereographic coordinate on the $\mathbb{P}^1$ fiber, and $Dt$ is a one-form on $\CZ$ that can be written as 
\begin{equation}
Dt \equiv dt+ p^+ -ip^3t+p^-t^2\,,
\end{equation}
where the vector $\vec{p}$ transforms as an SU(2) connection under SU(2) frame rotations. Let $\CU_i$ be an open covering of $\CZ$. On each $\CU_i$ one can define a one-form $\mathcal{X}^{[i]}$ which is patch-wise defined as
\begin{equation}\label{DefX}
\mathcal{X}^{[i]} = 2 e^{\Phi^{[i]}} \frac{Dt}{t}\,,
\end{equation}
for some complex-valued function $\Phi^{[i]}$, called the \emph{contact potential}. At the perturbative level, the contact potential is independent of $t$ and related to the four-dimensional dilaton as 
\begin{equation}\label{rpert}
e^{\Phi_{\rm pert}} \equiv r_{\rm pert} = \frac14 \mathcal{R}^2 K - c\,,
\end{equation} 
where $\log K$ is the K\"ahler potential of the special K\"ahler manifold embedded in $\cM$ via the c-map~\cite{Cecotti:1988qn,Ferrara:1989ik}. The one-loop corrections is encoded in the parameter $c$ given by \cite{RoblesLlana:2006ez}
\begin{equation}
c=-\frac{\chi_{X_3}}{192\pi} \,,
\end{equation} 
with $\chi_{X_3}$ the Euler characteristic of the threefold. Moreover, $\mathcal{R}$ is an auxiliary field that, under mirror symmetry, gets mapped to the ten-dimensional string coupling of Type IIB string theory. 

The metric \eqref{eq:metrictwistor} can then be obtained from a K\"ahler potential on $\CZ$ that is patch-wise defined in terms of the contact potential as 
\begin{equation}
K^{[i]} = \log\left(2\frac{1+t\bar{t}}{|t|}\right) +\text{Re}\,\Phi^{[i]}(x^\mu, t)\,. 
\end{equation}
Importantly, the contact structure can locally be trivialized to take the form 
\begin{equation}
\mathcal{X}^{[i]} \equiv {\rm d}\tilde{\alpha}^{[i]} + \xi^{[i]} {\rm d}\tilde{\xi}^{[i]}\,,
\end{equation} 
where $(\tilde{\alpha}, \xi^\Lambda, \tilde{\xi}_\Lambda)$ are referred to as \emph{holomorphic Darboux coordiantes}. Through their dependence on the coordinates $x^\mu$ of $\cM$ and $t$, these Darboux coordinates are the central object to infer the quantum corrections to the metric on $\cM$. When viewed as functions of $t$ for a fixed point $x^\mu \in \cM$, the function $\xi$ is allowed to have a simple pole at $t=0$ and $t=\infty$ whereas $\tilde{\xi}$ and $\tilde{\alpha}$ can have logarithmic singularities at these points. Suppose that $\mathcal{U}_+$ denotes the patch that contains the north pole of $\mathbb{P}^1$, i.e. $t=0$. The Darboux coordinates around $t=0$ have the expansion
\begin{equation}
\begin{aligned}
\xi^{[+]}&= \xi^{[+]}_{-1} t^{-1} + \xi^{[+]}_0 +\xi_1^{[+]} t +\mathcal{O}(t^2)\,,\\
\tilde{\xi}^{[+]} &= c_{\tilde\xi} \log t +  \tilde\xi^{[+]}_0 +\tilde\xi_1^{[+]} t +\mathcal{O}(t^2)\,,\\
\tilde{\alpha}^{[+]} &= c_{\tilde\alpha} \log t + \tilde\alpha^{[+]}_0 +\tilde\alpha_1^{[+]} t +\mathcal{O}(t^2)\,,\\
\Phi^{[+]} &= \phi_0^{[+]} + \phi_1^{[+]} t +\mathcal{O}(t^2)\,. 
\end{aligned}
\end{equation} 
For both the perturbative and the instanton corrected metric, there exist integral equations determining the Darboux coordinates explicitly in terms of $(x^\mu, t)$. We do not give their detailed form here, but refer to \cite{Alexandrov:2008gh,Alexandrov:2011va,Alexandrov:2023hiv} for details. 

Instead, we now focus on the universal hypermultiplet for which the moduli space metric is given in terms of the Przanowski description of self-dual Einstein spaces \cite{Przanowski:1983xpa}. The key insight is that for such self-dual Einstein such as $\cM$ the metric can be described locally by one real function $h$ which itself is a function of two complex coordinates $z^\alpha$, $\alpha=1,2$. This function needs to satisfy
\begin{equation}
\text{Prz}(h) \equiv h_{1\bar{1}} h_{2\bar{2}} - h_{1\bar{2}}h_{\bar{1}2}+(2h_{1\bar{1}}-h_1h_{\bar{1}}) e^h = 0\,. 
\end{equation}
The solution to this equation then gives a self-dual Einstein metric on an open subset $\mathcal{U}\subset \cM$ 
\begin{equation}\label{Przanowskimetric}
{\rm d}s^2_{\cM} = -\frac{6}{\Lambda} \left(h_{\alpha \bar{\beta}} {\rm d}z^\alpha {\rm d} z^{\bar\beta} +2e^h |{\rm d} z^2|^2\right)\,.
\end{equation}
This provides a relatively simple way to describe the metric on the universal hypermultiplet moduli space $\cM$, provided the function $h$ is known. Using the twistorial description of $\cM$ reviewed above, \cite{Alexandrov:2009vj} translates the metric \eqref{eq:metrictwistor} into the Przanowski metric in the following way. First one rewrites \eqref{eq:metrictwistor} as
\begin{equation} 
 {\rm d}s^2_{\cM} = \frac{12}{\Lambda} \left({\rm d}s^2_{\CZ} - e^{-2K} |\mathcal{X}|^2\right)\,,
\end{equation} 
and restricts to a real codimension-2 submanifold $\mathcal{C}$ given by the vanishing of some holomoprhic function $\mathcal{C}(u^i)=0$. Then the first part of the metric gives a K\"ahler metric with K\"ahler potential $K_{\mathcal{C}}$ that is the restriction of $K$ to $\mathcal{C}$. For the second part, one can assume that 
\begin{equation}
\CX|_{\mathcal{C}} = e^{F_{\mathcal{C}}(z^1,z^2)} {\rm d}z^2\,,
\end{equation}
with $\CX$ the one-form defined in \eqref{DefX} and $F_{\mathcal{C}}$ some holomorphic function defined on $\mathcal{C}$. The Przanowski metric is then reproduced for
\begin{equation}
 h= -2K_{\mathcal{C}} + F_{\mathcal{C}} +\bar{F}_{C}\,.
\end{equation}
The Przanowski potential $h$ is thus identified as the K\"ahler potential $K_{\mathcal{C}}$ subject to a K\"ahler transform given by $F_{\mathcal{C}}$. A convenient choice for the submanifold $\mathcal{C}$ is given by $t(x^\mu)=0$. As discussed in \cite{Alexandrov:2009vj} on this sub-locus $h$ can be expressed as 
\begin{equation}\label{hend}
h = -2\phi^{[+]}_0 + 2\log(\xi_{-1}^{[+]}/2)  = - 2\log\frac{2\tau_2 r}{\mathcal{R}}\,.
\end{equation} 
In terms of the Darboux coordinates, the coordinates $z^1$ and $z^2$ appearing in the definition \eqref{Przanowskimetric} of the Przanowski metric are given by 
\begin{equation}\label{eq:defz1z2}
z^1 = \frac{\ii}{2} \alpha_0^{[+]} -2c \log\xi_{-1}^{[+]} \,,\qquad z^2 = \frac{\ii}{2}\tilde{\xi}_0^{[+]} \,,
\end{equation} 
where the Fourier coefficients of the Darboux coordinates take the form
\begin{equation}
\begin{aligned}
 \xi_{-1}^{[+]} =&\; \mathcal{R}\,,\\
 \tilde{\xi}_0^{[+]} =& \; \tilde{\zeta} +\tau \zeta + \frac{1}{8\pi^2} \sum_\gamma \sigma_\gamma \bar{\Omega}_\gamma V_{\gamma}\cJ _\gamma^{(0)} -  \frac{1}{8\pi^2} \sum_{\boldsymbol{\gamma}} \bar{\Omega}_\gamma k \cL _{\boldsymbol{\gamma}}^{(0)} \,,\\
 \alpha^{[+]}_0 =& -\frac{1}{2} \left(\sigma + \zeta \tilde{\zeta} +\tau  \zeta^2\right) + 2\ii (r+c) \\ &+ \underbrace{\frac{\ii}{16\pi^3} \sum_\gamma \sigma_\gamma \bar{\Omega}_\gamma \left[(1+2\pi \ii\zeta V_{\gamma})\cJ _\gamma^{(0)}+2\pi \ii \mathcal{R}Z_\gamma \cJ_\gamma^{(1)} \right]}_{\text{M2-brane instantons}}\\
 &+\underbrace{\frac{\ii}{16\pi^3} \sum_{\boldsymbol{\gamma}} \bar{\Omega}_\gamma\left[\cI_{\boldsymbol{\gamma}}^{(0)} - 2\pi \ii k\zeta  \cL _{\boldsymbol{\gamma}}^{(0)} - 2\pi \ii k\mathcal{R} \mathscr{L}_{\boldsymbol{\gamma}}^{(1)}\right]}_{\text{M5-brane instantons}}\,,
\end{aligned}
\end{equation}
where the instanton contributions are split into M2- and M5-brane contributions which are expressed in terms of a shorthand notation that we review below. Let us notice that we can write $\alpha_0^{[+]}$ as 
\begin{equation}\begin{aligned}\label{alpha0+:simple}
 \alpha_0^{[+]} =& -\frac12 \sigma +2\ii (r+c) + \zeta \tilde{\xi}_0^{[+]} + \frac{\ii}{16\pi^3} \sum_\gamma \sigma_\gamma \bar{\Omega}_\gamma \left[\cJ _\gamma^{(0)}+2\pi \ii \mathcal{R}Z_\gamma \cJ_\gamma^{(1)} \right]\\ 
 &+\frac{\ii}{16\pi^3} \sum_{\boldsymbol{\gamma}} \bar{\Omega}_\gamma\left[\cI_{\boldsymbol{\gamma}}^{(0)} -2\pi \ii k\mathcal{R} \mathscr{L}_{\boldsymbol{\gamma}}^{(1)}\right]\,,
\end{aligned} \end{equation}  
where we used the definition of $\tilde{\xi}_0^{[+]}$. 

The description of the M2-brane instantons is considerably simpler since they map to D-instantons in the hypermultiplet sector of Type IIA on the same threefold, which have been considered already in \cite{RoblesLlana:2006ez,RoblesLlana:2006is,RoblesLlana:2007ae, Alexandrov:2008gh,Alexandrov:2009zh,Alexandrov:2012au,Alexandrov:2017qhn}. These depend on the instanton charge vector $\gamma=(p,q)$, which determines an element of $H_3(X_3,\mathbb{Z})$ corresponding to the three-cycle class wrapped by the M2-brane. To a given charge $\gamma$, one can associate the rational, generalized Donaldson--Thomas (DT) invariant
\begin{equation}
\bar{\Omega}_\gamma = \sum_{d|\gamma}\frac{1}{d^2}\Omega_{\gamma/d} \,,
\end{equation}
where $\Omega_\gamma$ are the generalized integer DT invariants that count the number of BPS instantons of a given charge. Furthermore $\sigma_\gamma$ is the quadratic refinement $\sigma_\gamma = (-1)^{qp}$. The instanton contribution then depends on the function $\cJ_\gamma$ defined as 
\begin{equation}
\cJ_\gamma = \mathscr{J}_\gamma \left[e^{-2\pi i(q\xi -p\tilde \xi)}\right]\,,
\end{equation}
where, for a function $H$, the integral transform $\mathscr{J}[H]$ is defined as 
\begin{equation}
\mathscr{J}_{\gamma}[H] (t)= \int_{C_i} \frac{dt'}{t'}\frac{t'+t}{t'-t} H \,.
\end{equation} 
Here $C_i$ is a contour on the $\mathbb{P}^1$ fiber of the twistor space and $t'$ the $\mathbb{P}^1$ coordinate along this contour. Furthermore, $(\xi, \tilde{\xi})$ are the Darboux coordinates introduced above. The $\mathcal{J}^{(n)}_\gamma$ are then the expansion coefficients of $\mathcal{J}(t)$ at $t=0$ and take the general form 
\begin{equation}
\mathcal{J}_\gamma^{(n)}=2\ii^{-|n|}(Z_\gamma/|\bar{Z}_\gamma|)^{n/2}e^{-2\pi i\Theta_\gamma} K_{|n|}(4\pi \mathcal{R}|Z_\gamma|) \,,
\end{equation} 
where $K_i$ is the modified Bessel function, $Z_\gamma$ the instanton central charge and 
\begin{equation}
\Theta_\gamma = q\zeta -p\tilde{\zeta}\,. 
\end{equation} 

The M5-brane instanton corrections are more involved. These instantons can be characterized by the charge vector $\boldsymbol{\gamma}=(k,p,q)$ where $k$ is the M5-brane instanton charge and $p,q$ induced M2-brane charge on the M5-brane instanton. The instanton function $\cI_{\boldsymbol{\gamma}}$ appearing in \eqref{alpha0+:simple} is given by 
\begin{equation}\label{Igammadef}
 \cI_{\boldsymbol{\gamma}} = \mathscr{J}_{\boldsymbol{\gamma}}\left[e^{2\pi i k\alpha_n}\Psi_{\hat{\boldsymbol{\gamma}}}\right]\,,
\end{equation} 
where $\alpha_n$ is defined in terms of the ratio $n=p/k$ of instanton charges as 
\begin{equation}
\alpha_n= - \frac12 \left(\widetilde{\alpha} + (\zeta-2n)\tilde{\zeta} \right)\,,
\end{equation}
with $\tilde{\alpha}$ the Darboux coordinate. The function $\Psi_{\hat{\boldsymbol{\gamma}}}$ depends on  $\zeta$ as detailed in \cite[Eq. (2.26)]{Alexandrov:2023hiv}. In general, the additional function appearing in the expression for the Darboux coordinates is given by 
\begin{equation}
\cL_{\boldsymbol{\gamma},\Lambda}^{(n)}=\cI_{\gamma,\Lambda}^{(n)}- F_{\Lambda \Sigma} \cI_{\boldsymbol{\gamma}}^{(n),\Sigma} \,\qquad \mathscr{L}_{\boldsymbol{\gamma},\Lambda} = z^\Lambda \cL_{\boldsymbol{\gamma},\Lambda}^{(n)}\,,
\end{equation}
where we recall that in the case of the universal hypermultiplet we have $F_{00}=-\tau=-(\tau_1+i\tau_2)$.

In the absence of the instanton correction, the only correction is perturbative in nature and is given by the parameter $c$. Dropping also this correction and comparing \eqref{eq:Sandxi} with \eqref{eq:defz1z2} we identify 
\begin{equation}
 S =- z^1 \,,\qquad \tilde{\xi} = -2\ii z^2 \,. 
\end{equation} 
Thus, as long as we restrict to the locus $\tilde{\zeta}=0$, the metric of the restricted moduli space is captured by the K\"ahler potential 
\begin{equation}
 K = - 2\log \frac{2r}{\mathcal{R}} \,. 
\end{equation}
Here, the real field $r$ is the full quantum volume of $X_3$ (or four-dimensional dilaton in case of Type IIA on $X_3$) which differs from $r_{\rm pert}$ in \eqref{rpert} through instanton corrections given by 
\begin{equation}
 r = \frac{\mathcal{R}^2}{4} K-c  +\frac{\ii \mathcal{R}}{32\pi^2} \sum_\gamma \sigma_\gamma \bar{\Omega}_\gamma \left(Z_\gamma \mathcal{J}_\gamma^{(1)} + \bar{Z}_\gamma \mathcal{J}_\gamma^{(-1)}\right) - \frac{\ii \mathcal{R}}{32\pi^2} \sum_{\boldsymbol{\gamma}} \bar{\Omega}_{\gamma} k \left(\mathscr{L}_{\boldsymbol{\gamma}}^{(1)}+ \bar{\mathscr{L}}_{\boldsymbol{\gamma}}^{(-1)}\right)\,. 
\end{equation} 
Comparing the instanton corrected version of $r$ with the definition of $S$ we conclude that the real part of $S$ is given by 
\begin{equation}\label{ReS}
 \text{Re}\,S= \frac{\mathcal{R}^2}{4}K + 2c \log(\mathcal{R}) + \frac{1}{32\pi^3}\text{Re}\left[\sum_\gamma \sigma_\gamma \bar{\Omega}_\gamma \mathcal{J}_\gamma^{(0)} + \sum_{\boldsymbol{\gamma}} \bar{\Omega}_\gamma \,\mathcal{I}_{\boldsymbol \gamma}^{(0)}\right]\,. 
\end{equation}
Notice that this agrees with the perturbative definition of the Przanowski coordinate \cite{Alexandrov:2006hx} up to the instanton corrections. 

\subsection{Corrected Domain Wall Solution}\label{sec:corrDW}
We now turn to the effect of the corrections to the hypermultiplet sector on the domain wall solution. We first discuss these corrections in general and then provide a qualitative study of the effect of the quantum corrections by focusing on the leading instanton approximation. 

\subsubsection{General analysis} 
From our previous discussion, we conclude that the complex scalar field running in the domain wall solution is still given by $S$, though its relation to the classical volume modulus $V\hat{=}\,r$ is corrected at the quantum level. Given the BPS domain wall equations \eqref{BPSequations}, we find 
\begin{equation}\label{eomsimple}
3  K^{S\bar{S}} K_{\bar{S}}  \partial_y a(y) = - \partial_y S(y)\,,
\end{equation} 
where we used 
\begin{equation}
 W(S,\bar{S}) = e^{K} \alpha \,,\qquad \partial_{\bar S} W(S,\bar{S}) = K_{\bar{S}} \,W(S,\bar{S})\,. 
\end{equation} 
We notice that all corrections to the domain wall solution arise due to corrections to the K\"ahler potential $K$. In particular the sources $J^{(a)}$ associated to the two orbifold planes are not affected by the corrections. Accordingly, the harmonic function $H(y)$, that determines the profile of the running solution, continues to be given by \eqref{Hofy}, and the relation between $S$ and $H(y)$ in \eqref{classicalsolution} is likewise unaffected by the corrections. We hence still have 
\begin{equation}\label{Sprofile}
 S = V_0 \left(c_0 + \frac{\sqrt{2}\alpha}{3} |y|\right)^3\,. 
\end{equation} 
In particular since $\partial_y S = 3 V_0 H^2(y) \partial_y H(y)$ the RHS of \eqref{eomsimple} is finite for any finite $y$ and vanishes for $y=y_{\rm max}$. Hence $\partial_y a$ is also finite unless the prefactor $K^{S\bar{S}} K_{\bar S}$ goes to zero faster for $y\to y_{\rm max}$ than the RHS of \eqref{eomsimple}. In the absence of instanton corrections, this latter case is indeed realized since classically we have 
\begin{equation}
K^{S \bar S}_{\rm class.} \sim -(S+\bar{S})^2 \,,\qquad K_{\bar{S}}^{\rm class} \sim -(S+\bar{S})^{-1}\,. 
\end{equation}
As a consequence, $\partial_y a$ cannot be finite as $y\rightarrow y_{\rm max}$ for the classical solution implying that $a(y\to y_{\rm max})\to -\infty$ and the warp factor $e^{2a} \to 0$ as expected for an end-of-the-world brane. \\

The situation is different at the quantum level, where the corrections to the K\"ahler potential have to be taken into account. Before turning to the actual quantum corrections to the hypermulitplet moduli space discussed in the previous section, let us first consider a simple toy model that illustrates the effect of quantum corrections on the domain wall solution. Therefore, suppose that the instanton corrections to the K\"ahler potential have the following simple form 
\begin{equation}\label{Ktoy}
e^{-K^{\rm toy} (S,\bar{S})}= (S+\bar{S}) + b \exp\left[-(S+\bar{S})\right]\,,
\end{equation} 
corresponding to a single instanton correction whose effect on the K\"ahler potential is encoded in the parameter $b$. For simplicity, we assume $b$ to be constant. Importantly, our toy K\"ahler potential is still a function of $(S+\bar{S})$ only. Since $K^{\rm toy}_{\bar{S}}=K^{\rm toy}_S$ we can rewrite \eqref{eomsimple}  as
\begin{equation}
 \partial_y a(y) = -\frac12 \partial_y \log K^{\rm toy}_S \,,
\end{equation}
which can be integrated to 
\begin{equation}\label{integratedequation}
2 a(y) = - \log K^{\rm toy}_S+\text{const}\,.
\end{equation}
Using the toy version of the K\"ahler potential \eqref{Ktoy} we find 
\begin{equation}
 K_S^{\rm toy} = \frac{1 - b \,e^{-(S+\bar{S})}}{S+\bar{S} + b e^{-(S+\bar{S})} }\,,
\end{equation}
which, for $0<b<1$, remains finite and non-zero in the limit $S\rightarrow 0$. For this range of the parameter $b$, the warp factor $a(y)$ given in \eqref{integratedequation} thus also remains finite at the invisible brane, implying that this brane is not an end-of-the-world brane at which space-time caps off. Thus, depending on the value (and in particular the sign) of the coefficient $b$ of the leading instanton corrections, the classical singularity at the end-of-the-world brane gets resolved into a brane with finite warp factor at the quantum level. Still, the above analysis relies on the fictitious toy K\"ahler potential in \eqref{Ktoy}. Given the explicit expressions for the instanton corrections to the universal hypermultiplet moduli space discussed in the previous section, we can now be more explicit and, in particular, determine the sign of the leading instanton correction. \\

To see the actual effect of the quantum corrections on the domain wall solution, we recall from the previous section that the K\"ahler potential for the field $S$ is given by 
\begin{equation}\label{Kahlerofr}
K = - 2\log\frac{2r}{\mathcal{R}}\,,
\end{equation}
where $r$ and $\mathcal{R}$ implicitly depend on $S$ as is clear by inverting \eqref{ReS}. Notice that the instanton corrections encoded in $\mathcal{I}_{\boldsymbol \gamma}^{(0)}$ also depend on $\mathcal{R}$, such that the relation \eqref{ReS} is difficult to invert. Thus, determining the $S$-dependence of the K\"ahler potential is similarly non-trivial. Here, we refrain from explicitly attempting to write down the K\"ahler potential as a function of $S$ but restrict ourselves to a qualitative analysis of the behavior of the K\"ahler potential for $S\rightarrow 0$. In particular, our goal is to determine whether the K\"ahler potential and its derivatives are finite as we approach $\text{Re}\,S\rightarrow 0$. As a first step we notice that  
\begin{equation}
  K_S = \left(\frac{\partial \mathcal{R}}{\partial S} \right)\partial_{\mathcal{R}} K\,,
\end{equation} 
and similarly for the second derivatives. The simplified BPS domain wall equation \eqref{eomsimple} then takes the form 
\begin{equation}\label{derivativeay}
3 \partial_y a(y) =  \frac{\mathcal{R} \tau_2}{2r}\left(\frac{\partial \mathcal{R}}{\partial S}\right)\partial_y S(y)\,. 
\end{equation} 
Thus, in order for the warp factor to diverge at $S=0$ (corresponding to the classical end-of-the-world brane), we either need $r\rightarrow 0$, $\mathcal{R}\rightarrow 0$, or $\frac{\partial \mathcal{R}}{\partial S} \rightarrow \infty$ for $S\rightarrow 0$. 

We can first consider this problem at the perturbative level where $r$ and $S$ are related via
\begin{equation}\label{Spert}
 \text{Re}\,S_{\rm pert}=\frac{\mathcal{R}^2}{4}\tau_2+ 2c \log(\mathcal{R}) \,,
\end{equation}
where we just dropped all instanton effects in \eqref{ReS}. We then have
\begin{equation}\label{derivativer}
  \frac{\partial \mathcal{R}}{\partial S_{\rm pert}} = \left(\frac{\partial S_{\rm pert}}{\partial \mathcal{R}}\right)^{-1} =  \left(\frac{\partial \left(\text{Re}\,S_{\rm pert}\right)}{\partial \mathcal{R}}\right)^{-1} = \left(\frac{\mathcal{R}\tau_2}{2}+\frac{2c}{\mathcal{R}}\right)^{-1}\,,
\end{equation} 
where in the second step we used that $\text{Im}\,S_{\rm pert}$ is independent of $\mathcal{R}$. For rigid Calabi--Yau threefolds, the Euler characteristic is positive such that $c<0$.\footnote{An earlier version of this paper had $c>0$, we thank Sergei Alexandrov for pointing this out to us.} Since the RHS of \eqref{derivativer} diverges for $\mathcal{R}^2=-4c/\tau_2$, by \eqref{derivativeay} a singularity for the warp factor of the domain wall solution remains at the perturbative level.\footnote{Notice that at the perturbative level $\mathcal{R}^2=-4c\tau_2$ corresponds to $r=-2c$ which is a singularity of the perturabtive metric on the universal hypermultiplet moduli space~\cite{Alexandrov:2009qq} which has to be resolved by NS5-brane instantons~\cite{Alexandrov:2009qq,Alexandrov:2014sya}.} Note, however, that due to the logarithmic correction to $\text{Re}\,S$, achieving $\text{Re}\,S=0$ requires turning on the axions in the universal hypermultiplet, which is inconsistent with the dilatonic domain wall solution. Thus, at the perturbative level, the BPS equations do not have a domain wall solution. This should, however, not be seen as evidence that no domain wall solution exists, but just that we cannot trust the perturbative result since in the regime of interest $\mathcal{R}\ll 1$ and, hence, the classical and one-loop terms in $\text{Re}\,S$ are of the same order.

Instead, in the strong coupling regime that we consider here, the dominant quantum correction comes from the aforementioned 5-brane instantons. These correct $\text{Re}\,S$ as a function of $\mathcal{R}$ as in \eqref{ReS}. We expect these corrections to ensure that the domain wall solution still exists at the quantum level in such a way that the warp factor $a(y)$ remains finite as we approach $y\to y_{\rm max}$. To that end, we need to ensure that $i)$ a domain wall solution realizing strong coupling ($\text{Re}\,S=0$) exists, and $ii)$ that the RHS of \eqref{derivativeay} remains finite for this domain wall for any $y\leq y_{\rm max}$. To achieve the first point, the instanton corrections need to correct $\text{Re}\,S$ in such a way that $\text{Re}\,S=0$ is realized for $\mathcal{R}>0$. Let us schematically write the instanton corrections in \eqref{ReS} as
\begin{equation}\label{eq:M5schematic}
\delta(\text{Re}\,S) \sim \sum_{n} \mathcal{A}_{n} e^{-4\pi n \mathcal{R}^2} \,,
\end{equation}
for some coefficients $\mathcal{A}_n$.
This correction needs to be negative in order to realize $\text{Re}\,S=0$ for finite $\mathcal{R}>0$. On the other hand, taking the schematic correction in \eqref{eq:M5schematic} into account when computing the derivative in \eqref{derivativer}, we find
\begin{equation}\label{derivativercorr}
 \frac{\partial\left(\text{Re}\,S_{\rm pert}+\delta(\text{Re}\,S)\right)}{\partial \mathcal{R}} = \frac{\mathcal{R}\tau_2}{2}+\frac{2c}{\mathcal{R}} - \sum_n (8\pi n \mathcal{R}) \mathcal{A}_{n} e^{-4\pi n \mathcal{R}^2}\,. 
\end{equation} 
If the correction term in the expression above is negative, the derivative vanishes for some $\mathcal{R}^2>-4c/\tau_2$. By \eqref{derivativeay} this implies that also the warp factor of the domain wall solution diverges at this point. On the other hand, if the corrections to the above derivative are positive, the singularity on the RHS of \eqref{derivativeay} is shifted to $\mathcal{R}^2<-4c/\tau_2$. Thus, in case $\mathcal{A}_n<0$, it is possible to reach the strong coupling point for the hidden gauge theory ($\text{Re}\,S=0$) before reaching the singularity in the warp factor and hence to obtain a well-defined domain wall solution realizing strong coupling ($\text{Re}\,S=0$) for which the warp factor remains finite. 

We thus need to extract the sign of the instanton corrections to determine whether the warp factor of the domain solution has a singularity as $\text{Re}\,S\rightarrow 0$. To that end, one would need to compute the entire instanton sum. However, this is out of reach, given the current computational techniques. We therefore restrict to the leading instanton and investigate whether the sign of this term corrects $\text{Re}\,S$ as a function of $r$ in such a way as to ensure that the warp factor of the domain wall solution is finite for all $y<y_{\rm max}$.

\subsubsection{Leading instanton approximation} 
To study the leading instanton contribution to $\text{Re}\,S$ as a function of $r$, it is instructive to first consider the instanton correction to $\text{Re}\,S$ in \eqref{ReS} in the weak coupling limit. In the following, we focus on the M5-brane instanton captured by the real part of the function $\mathcal{I}_{\boldsymbol{\gamma}}^{(0)}$.  The weak coupling behavior of $\mathcal{I}_{\boldsymbol{\gamma}}^{(0)}$ has been analyzed in \cite{Alexandrov:2023hiv}. More precisely, the weak coupling limit is defined as the scaling limit
\begin{equation}\label{weakcouplinglimit}
r = r_0\, g_s^{-2} \,,\quad \sigma \sim \sigma_0 \, g_s^{-2} \,\quad \zeta \sim \zeta_0 \, g_s^{-1}\,,\quad  \tilde{\zeta}  \sim\tilde{\zeta}_0\, g_s^{-1} \,,\qquad  g_s\rightarrow 0 \,,
\end{equation} 
for some fixed $(r_0,\sigma_0,\zeta_0,\tilde{\zeta}_0)$. Since in \cite{Alexandrov:2023hiv} the corrections due to NS5-brane instantons in Type IIA compactified on IIA, the parameter $g_s$ is identified as the ten-dimensional string coupling of Type IIA string theory. Here, we are interested in M5-brane instanton corrections to the hypermultiplet moduli space of M-theory on $X_3$ such that we treat $g_s$ as a scaling parameter without a direct physical interpretation. In the limit \eqref{weakcouplinglimit}, the function $\mathcal{I}_{\boldsymbol{\gamma}}^{(0)}$ for $k>0$ can be evaluated to be \cite{Alexandrov:2023hiv} 
\begin{equation}\label{Iapprox}
\mathcal{I}_{\boldsymbol{\gamma}}^{(0)} \simeq -\frac{k}{p^0} \frac{\xi^0(t_0)}{t_0^{1+8\pi k c}} \frac{e^{-S_{\boldsymbol{\gamma}}}}{\sqrt{\ii k \mathcal{S}_0''(t_0)}}\,,
\end{equation}
such that without the perturbative and M2-brane instanton corrections, $\text{Re}\,S$ is given by 
\begin{equation}\label{ReSleading}
\text{Re}\,S \rightarrow r - \frac{1}{32\pi^3} \sum_{{\boldsymbol{\gamma}}} \text{Re}\left[ \frac{\bar{\Omega}_\gamma k}{p_0}\frac{\xi^0(t_0)}{t_0^{1+8\pi kc}} \frac{e^{-S_{\rm \gamma}}}{\sqrt{\ii k \cS_0''(t_0)}} \right]\,.
\end{equation} 
The exponential term in \eqref{Igammadef}  can be interpreted as an effective instanton action $\mathcal{S}_{\boldsymbol{\gamma}(t)}$ where $t$ is the coordinate on the $\mathbb{P}^1$-fiber of the twistor space $\mathcal{Z}$. In the limit \eqref{weakcouplinglimit}, \cite{Alexandrov:2023hiv} expands $\mathcal{S}_{\boldsymbol{\gamma}}(t) = -4\ii c\log t + \sum_{l\geq 0}\mathcal{S}_{\boldsymbol{\gamma,l}}$ where the $l$-th term scales as $g_s^{l-2}$. The exact expressions for the leading $\mathcal{S}_{\boldsymbol{\gamma},l}$ are not relevant to our discussion and can be found in \cite[Eq. (5.16)]{Alexandrov:2023hiv}. The saddle point approximation is given by evaluating the effective action at the point $t_0$ corresponding to the solution of $\mathcal{S}_{\boldsymbol{\gamma},0}'\equiv \partial_t \mathcal{S}_{\boldsymbol{\gamma},0}=0$ such that the instanton action in \eqref{Iapprox} in the limit \eqref{weakcouplinglimit} is given by \cite{Alexandrov:2023hiv}
\begin{equation}
 S_{\boldsymbol \gamma} = 2\pi i k\left(\mathcal{S}_{\boldsymbol{\gamma},0}+\mathcal{S}_{\boldsymbol{\gamma},0}+\mathcal{S}_{\boldsymbol{\gamma},1}+\mathcal{S}_{\boldsymbol{\gamma},2}-\frac12 \frac{(\mathcal{S}_{\boldsymbol{\gamma},1}')^2}{\mathcal{S}_{\boldsymbol{\gamma},0}''}\right)\,. 
\end{equation} 
In terms of the coordinates on the hypermultiplet moduli space, the saddle point is given by $t_0=\frac{\zeta}{\mathcal{R}}$ which hence does not scale with $g_s$. On top of the limit \eqref{weakcouplinglimit} one can choose the limit of small Ramond-Ramond(RR)-fields (in the language of Type II string theory) by choosing $0<\zeta_0\ll 1$. Since we are interested in the sub-locus of the universal hypermultiplet sector along which $z^2=0$, this also requires tuning $\tilde{\zeta}$ accordingly to another small value. Taking such a double-scaling limit, \cite{Alexandrov:2023hiv} evaluates 
\begin{equation}
\mathcal{S}_0''(t_0) = - i \mathcal{R}^2 \tau_2\,,
\end{equation}
such that the argument of the square root in \eqref{Iapprox} is real and positive. Moreover, the instanton action in this limit is then given by
\begin{equation}
S_{\boldsymbol{\gamma}} = 4\pi k r + \pi \ii k\left[\sigma + (\zeta-2n) \tilde{\zeta}+ \tau (\zeta-n)^2 \right] \,,
\end{equation} 
where $n=p/k$ (see also \cite{Witten:1996hc,Belov:2006jd}). \\

We now apply the above discussion to the leading instanton. This will enable us to assess the qualitative effect of the instanton correction on the domain wall solution. As discussed previously, the sign of this correction is particularly interesting. To determine this sign, we again restrict to the locus $z^2=0$, $\text{Im}\,S=0$ in the universal hypermultiplet moduli space.  As we will argue below, to determine the sign of the correction to $\text{Re}\,S$ it is sufficient to consider the small coupling limit \eqref{weakcouplinglimit}. In this limit, we can apply the general discussion above to the special case of the leading instanton with charge  $\boldsymbol{\gamma}_0=(1,0,\vec{0})$. Notice that this choice of charge vector corresponds to the standard D-brane charge vector $\gamma_0=(1,\vec{0})$ such that in the above expression $k=p^0=1$. In this approximation the corrections to $\text{Re}\,S$ take the simple form 
\begin{equation}\label{ReSoneinst.}
\text{Re}\,S \simeq r +c\log (r+c)- \frac{1}{32\pi^3} \text{Re}\left[ \bar{\Omega}_{\gamma_0}\frac{\xi^0(t_0)}{t_0^{1+8\pi c}} \frac{e^{-S_{\rm \gamma_0}}}{\sqrt{\ii \cS_0''(t_0)}} \right]\,,
\end{equation} 
where we used the saddle point approximation \eqref{Iapprox} for $\mathcal{I}_{\boldsymbol{\gamma}}^{(0)}$. For the leading instanton, the saddle point approximation in the weak-coupling limit is given by 
\begin{equation}
S_{\boldsymbol{\gamma}_0} = 4\pi r + \pi \ii\left[\sigma + \zeta(\tilde{\zeta}+ \tau \zeta) \right] \,.
\end{equation} 
We recognize the second term in the imaginary part of $S_{\boldsymbol{\gamma}_0}$  as being proportional to the classical expression for $z^2$, which we set to zero on the locus in the universal hypermultiplet moduli space that we consider here. We further set $\sigma_0=0$ which corresponds to the locus $\text{Im}\,S=0$ such that the instanton action is real. Furthermore, since the square root appearing in the denominator of \eqref{ReSoneinst.} is real and positive and we choose $0<\zeta_0\ll 1$, the sign of the correction in \eqref{ReSoneinst.} only depends on the sign of $\bar{\Omega}_{\gamma_0}$. 

The sign of $\bar{\Omega}_{\gamma_0}$ can be determined using Type IIA string theory compactified on the same Calabi--Yau threefold. In this case, the charge $\gamma_0$ corresponds to the charge of a single D6-brane wrapping the entire Calabi--Yau threefold. For this simple charge vector, the invariant ${\bar \Omega}_{\gamma_0}$ is expected to coincide with the BPS index of the brane. The D6-brane wrapping the entire Calabi--Yau is the massless state arising at the principal component of the discriminant locus of the quantum K\"ahler moduli space of Type IIA compactified on $X_3$. This singularity corresponds to a conifold singularity, implying that a single hypermultiplet becomes massless along this singular locus, which we identify as the wrapped D6-brane. As a consequence, the BPS index for the single-wrapped D6-brane is $\bar{\Omega}_{\gamma_0}=1>0$. 

Putting things together, we conclude that for the first instanton contribution, the term in the brackets in \eqref{ReSoneinst.} in the limit \eqref{weakcouplinglimit} is real and positive. Thus, the leading instanton correction to $\text{Re}\,S$ in terms of $r$ is negative. In the above discussion, we effectively considered the locus $\text{Im}\,S=0$ in the universal hypermultiplet sector as parametrized by $g_s$ as in \eqref{weakcouplinglimit}. The result of this consideration is that the leading instanton correction in the limit $g_s\rightarrow 0$ to $\text{Re}\,S$ is negative since the term in \eqref{ReSoneinst.}, of which the real part is taken, is real and negative. In fact, as we go to finite $g_s$, the sign of the instanton correction should not change since this would imply the existence of a finite distance point in the universal hypermultiplet moduli space at which the instanton contribution vanishes. This should, however, only happen at special loci in the moduli space, of which there are none in the universal hypermultiplet moduli space. This argument then implies that the leading instanton correction to $\text{Re}\,S$ has a negative sign also away from the weak coupling limit for finite $r$. \\ 

With this preparation, we can return to the domain wall solution. Recall from our discussion below \eqref{eq:M5schematic} that the leading instanton contribution to $\text{Re}\,S$ is the relevant correction that determines whether the singularity of the warp factor in the domain solution persists at the quantum level. In the light of \eqref{derivativeay} this is ensured if $i)$ the corrections ensure that $\text{Re}\,S=0$ can be satisfied for some $\mathcal{R}>0$, and $ii)$ the derivative $\partial \mathcal{R}/\partial S$ is finite everywhere. As discussed below \eqref{eq:M5schematic} both points can be satisfied if the instanton correction to $\text{Re}\,S$ is negative. As we showed above, this is indeed the case as far as the leading instanton correction is concerned.  We hence conclude that 
\begin{equation}\label{partialay}
\partial_y a(y) = -\frac{1}{3 K_{\bar{S}}} K_{S \bar{S}} \partial_y S(y) = {\rm finite}\,\quad \text{as} \quad y\to y_{\rm max}\,,
\end{equation} 
and since the derivative of $a(y)$ is finite for finite $y$ also $a(y\to y_{\rm max})$ itself has to be finite. Thus, the instanton corrections to the hypermultiplet metric ensure that, at the quantum level, there is no divergence of the warp factor of the BPS domain wall solution. 

Our results indicate that the non-perturbative corrections resolve the classical end-of-the-world brane appearing at the strong coupling singularity for the hidden $E_8$ domain wall into a domain wall with finite curvature. We can now translate this result back to the original duality frame corresponding to the heterotic string compactified on a Calabi--Yau threefold discussed in section~\ref{sec:perthet}. The absence of the end-of-the-world brane in the M-theory domain wall solution implies that also the moduli space of the effective $\cN=1$ theory does not end at the strong coupling singularity. In particular, the spacetime does not end at $2\pi \rho = y_{\rm max}$, implying that larger values for $\rho$ can, in principle, be considered. Via \eqref{identification}, this means that also the field space of the heterotic string extends beyond $\text{Re}\,T_{\rm max}$. This should be contrasted to the six-dimensional cousin of the strong coupling singularity discussed in section~\ref{sec:6dcomment} for which the tensor branch cannot be extended beyond the locus associated with the strong coupling singularity. 

To see that in the four-dimensional theory, the field space extends beyond the classical boundary, it was crucial to consider the instanton corrections to the effective $\cN=1$ four-dimensional theory. In the effective four-dimensional theory, the relevant instantons are, however, not BPS and, as such, difficult to compute. Similarly, higher-order perturbative corrections are \emph{a priori} not necessarily vanishing. This difficulty has been circumvented in our approach by considering the 5d Horava--Witten origin of the four-dimensional theory in which these instanton effects are inherited from BPS instantons correcting the hypermultiplet sector of the 5d $\cN=1$ parent theory which furthermore, does not receive higher-order perturbative corrections.  Both perturbative and non-perturbative corrections can be explicitly computed, and their effect on the domain wall solution of the 5d $\cN=1$ theory revealed that the strong coupling singularity for the unbroken $E_8$ gauge group of the heterotic string is less severe than the classical analysis suggests.

\section{Conclusions}\label{sec:concl}
In this paper, we considered the fate of strong coupling singularities in the scalar field space of 4d $\cN=1$ effective theories of gravity at the quantum level. Specifically we focused on strong coupling singularities for the unbroken $E_8$ gauge group of the perturbative heterotic string compactified on Calabi--Yau threefolds with standard embedding. In this setup, the strong coupling singularities arise due to threshold corrections to the heterotic gauge couplings, which are hence invisible to the tree-level heterotic string. 

To investigate the effect of quantum corrections on the strong coupling singularities, we used the duality of the $E_8\times E_8$ heterotic string to M-theory on an interval with 9-branes located at the orbifold fixed points. If compactified on an additional Calabi--Yau threefold, the 9-branes carry an induced 5-brane charge. This charge is a source for the volume of the Calabi--Yau, which thus varies over the interval. The resulting setup can be described as a five-dimensional dilatonic domain wall, with the volume modulus playing the role of the dilaton. If the length of the M-theory interval is tuned to a specific critical value, we showed that the classical BPS domain wall solution has an end-of-the-world brane corresponding to a strong coupling limit for the gauge theory on the hidden Horava--Witten 9-brane accompanied by a divergence of the warp factor at the location of the brane. We interpreted the appearance of this end-of-the-world brane as dual to the strong coupling singularity of the perturbative heterotic string, thus illustrating that the strong coupling singularity arises in a gauge theory sector that cannot be decoupled from gravity. The warp factor of the domain wall solution diverges at the location of the hidden brane. For the classical domain wall solution, it is thus impossible to increase the size of the interval beyond a critical value, indicating that the strong coupling singularity serves as a boundary for the classical field space of the 4d $\cN=1$ theory. 

We have, however, shown that this conclusion does not hold at the quantum level. Such corrections to the BPS domain wall equations are captured by corrections to the hypermultiplet sector of the 5d $\cN=1$ parent theory. The instantons correcting the hypermultiplet sector of this parent theory are BPS, and perturbative corrections do not exist beyond the first order (one-loop in the analog 4d $\cN=2$ string theory). Thus, even though the instantons are non-BPS from the 4d $\cN=1$ perspective, non-perturbative corrections can be explicitly calculated due to their BPS origin in the 5d $\cN=1$ parent theory. Using the recent results of \cite{Alexandrov:2023hiv}, we argued that these corrections regulate the divergence of the warp factor at the location of the 9-brane and, hence, effectively resolve the end-of-the-world brane into a brane with finite curvature, which, however, remains strongly coupled. This led to the main conclusion of the paper: The classical strong coupling boundary of the scalar field space of the perturbative heterotic string gets resolved by instanton effects, indicating that, at the quantum level, this field space extends beyond the strong coupling regime.

Our results are thus an example of how the corrections to the hypermultiplet sector significantly alter the properties of the theory compared to the classical expectation. Another example of this kind was discussed in \cite{Marchesano:2019ifh, Baume:2019sry,Alvarez-Garcia:2021pxo}. There, it was shown that the non-perturbative quantum corrections to the hypermultiplet metric obstruct certain infinite distance limits of the classical hypermultiplet moduli space, implying that the asymptotic regimes of the classical moduli space differ significantly from those of the quantum moduli space. In particular, only the asymptotics of the fully quantum corrected moduli space satisfy the emergent string conjecture~\cite{Lee:2019oct} as shown in~\cite{Baume:2019sry, Alvarez-Garcia:2021pxo}. Together with these analyses, the results obtained in this paper highlight the importance of studying the role of quantum corrections on the geometry of field spaces in gravitational theories. \\

There are several open questions that we will leave for future work. First, we did not attempt to identify the actual description of the strong coupling phase that extends the quasi-moduli space beyond the classical strong coupling boundary. To approach this problem, it is interesting to recall that for open string gauge theories, the confined phase of the gauge theory can be described within string theory as in~\cite{Atiyah:2000zz, Acharya:2000gb}. In this case, the phase transition to the confined phase corresponds to a geometric transition in which D6-branes of Type IIA string theory are traded for RR-flux. We expect an analog transition for our case, but this time for the perturbative heterotic gauge theory, which is explicitly coupled to gravity. In analogy to the open string transition discussed in~\cite{Atiyah:2000zz, Acharya:2000gb}, we expect the gravitational version to replace the strongly coupled 9-brane by some flux. Since the warp factor of the domain wall solution remains finite at strong coupling, also the curvature is finite but still non-zero. It is therefore suggestive that in the strongly coupled phase of the 4d $\cN=1$ theory, the original manifold on which M-theory is compactified, $X_3\times S^1/\mathbb{Z}_2$, gets replaced by a 7-manifold with non-zero curvature, possibly together with some additional flux. Thus, in this picture, the strongly coupled 9-brane is effectively replaced by the curvature of the 7-manifold, which could be viewed as a gravitational analog of replacing the D6-branes of Type IIA with RR 2-form flux. Explicitly working out such a description for the confined phase in our gravitational setting is an exciting direction for future research. 

Moreover, we did not address the effect of the potential for the scalar fields in the chiral multiplets and effectively treated the scalar field space as a quasi-moduli space. On the other hand, for the heterotic string compactified on Calabi--Yau threefolds, a non-perturbative superpotential can be generated through gaugino condensation. For the simplest case of the heterotic string on $T^6/\mathbb{Z}_3$ with standard embedding, the non-perturbative superpotential and scalar potential have been analyzed in~\cite{Cvetic:1991qm} and more recently, e.g., in \cite{Gonzalo:2018guu,Leedom:2022zdm,Cribiori:2023sch}. In the vicinity of the strong coupling singularity, the scalar potential becomes of the order of the heterotic string scale and thus of the order of the quantum gravity cutoff --- again due to the threshold corrections. If one trusted this scalar potential even beyond the strong coupling singularity, the scalar potential would grow parametrically larger than the quantum gravity cutoff, signaling that the original EFT description breaks down~\cite{Hebecker:2018vxz,Scalisi:2018eaz,vandeHeisteeg:2023uxj}. This is not surprising since we need to change the effective description of the theory in the strongly coupled phase. It would be interesting to see whether the scalar potential, when computed within the correct low energy theory of the strongly coupled phase, gets regulated similarly to the strong coupling singularity. To compute the scalar potential, it is important to calculate the explicit form of the K\"ahler potential in the strong coupling regime. We recall that, in this work, we did not directly compute the corrections to the K\"ahler potential of the effective 4d $\cN=1$ theory. Instead, we computed the corrections to the domain wall solution of a 5d $\cN=1$ theory that reduces to the 4d $\cN=1$ theory. It would be interesting to see how these effects translate to corrections of the 4d $\cN=1$ K\"ahler potential and how the resolution of the strong coupling singularity is reflected in the corrected K\"ahler potential. We leave this direction for future research. 

The above points highlight that the results obtained in this paper open various new lines of investigation. In addition, it would be relevant to perform an analysis similar to what was done in this work for more general heterotic compactifications with non-standard embedding of the gauge bundles and their M-theory duals. Moreover, we restricted our analysis of the non-perturbative instanton corrections to the domain wall solutions to the leading instanton approximation, which allowed us to make qualitative statements about the behavior of the warp factor at strong coupling. A more thorough, quantitative analysis of this effect could reveal important information necessary to answer the above open questions.  

\subsubsection*{Acknowledgements} We thank Cumrun Vafa for helpful discussions and suggestions and Sergei Alexandrov for very useful comments on an earlier version of this paper. MC is supported by the DOE Award (HEP) DE-SC0013528, the Simons Foundation Collaboration grant \#724069, Slovenian Research Agency (ARRS No. P1-0306), and Fay R. and Eugene L. Langberg Endowed Chair funds. MC would like to thank the Harvard Swampland Initiative for hospitality, where this work was initiated. MW is supported by a grant from the Simons Foundation (602883,CV), the DellaPietra Foundation, and the NSF grant PHY-2013858.

\bibliography{papers_Max}
\bibliographystyle{JHEP}

\end{document}